\documentclass[10pt,journal,compsoc]{IEEEtran}

\usepackage{booktabs}
\usepackage{amsmath}
\usepackage{listings}
\usepackage{hyperref}
\usepackage{cleveref}
\usepackage{algorithm}
\usepackage{algorithmicx}
\usepackage[noend]{algpseudocode}
\usepackage{syntax}
\usepackage{subcaption}
\usepackage{multirow}
\usepackage{graphicx}
\usepackage{xcolor}
\usepackage{fancyvrb}
\usepackage{ragged2e}
\usepackage{array}
\usepackage{balance}
\usepackage{mathtools} 
\usepackage{amsmath}

\newtheorem{definition}{Definition}

\newcommand{\name}{DiffSearch}

\newcommand{\code}[1]{\texttt{\small#1}}
\newcommand{\scode}[1]{\texttt{\footnotesize#1}}

\ifCLASSOPTIONcompsoc
  \usepackage[nocompress]{cite}
\else
  \usepackage{cite}
\fi

\begin{document}

\title{\name{}: A Scalable and Precise\\Search Engine for Code Changes}

\author{Luca Di Grazia,
Paul Bredl, Michael Pradel% <-this % stops a space

	\thanks{All authors are with the Department
of Computer Science, University of Stuttgart, Germany. Email: luca.di-grazia@iste.uni-stuttgart.de, paulbredl@gmx.de, michael@binaervarianz.de
		}% <-this % stops an unwanted space

}

\markboth{IEEE Transactions on Software Engineering}%
{Di Grazia et al.: DiffSearch}

\IEEEtitleabstractindextext{%
\begin{abstract}
The source code of successful projects is evolving all the time, resulting in hundreds of thousands of code changes stored in source code repositories.
This wealth of data can be useful, e.g., to find changes similar to a planned code change or examples of recurring code improvements.
This paper presents \name{}, a search engine that, given a query that describes a code change, returns a set of changes that match the query.
The approach is enabled by three key contributions.
First, we present a query language that extends the underlying programming language with wildcards and placeholders, providing an intuitive way of formulating queries that is easy to adapt to different programming languages.
Second, to ensure scalability, the approach indexes code changes in a one-time preprocessing step, mapping them into a feature space, and then performs an efficient search in the feature space for each query.
Third, to guarantee precision, i.e., that any returned code change indeed matches the given query, we present a tree-based matching algorithm that checks whether a query can be expanded to a concrete code change.
We present implementations for Java, JavaScript, and Python, and show that
the approach responds within seconds to queries across one million code changes,
has a recall of 80.7\% for Java, 89.6\% for Python, and 90.4\% for JavaScript, enables users to find relevant code changes more effectively than a regular expression-based search and GitHub's search feature,
and is helpful for gathering a large-scale dataset of real-world bug fixes.

\end{abstract}

}

\maketitle

\IEEEraisesectionheading{\section{Introduction}\label{sec:intro}}

\IEEEPARstart{H}{undreds} of thousands of code changes are stored in the version histories of code repositories.
To benefit from this immense source of knowledge, practitioners and researchers often want to search for specific kinds of code changes.
For example, developers may want to search through their own repositories to find again a code change performed in the past, or search for commits that introduce a specific kind of problem.
Developers may also want to search through changes in repositories by others, e.g., to understand how code gets migrated from one API to another, or to retrieve examples of common refactorings for educational purposes.
A question on Stack Overflow on how to systematically search through code changes\footnote{\url{https://stackoverflow.com/questions/2928584/how-to-grep-search-committed-code-in-the-git-history}} has received over half a million views, showing that practitioners are interested in finding changes from the past.

Besides practitioners, researchers also commonly search for specific kinds of code changes.
For example, a researcher evaluating a bug finding tool~\cite{ase2018-study} or a program repair tool~\cite{cacm2019-program-repair,DBLP:conf/icse/TanYYMR17,motwani2020quality} may be interested in examples of specific kinds of bug fixes.
Likewise, researchers working on machine learning models that predict when and where to apply specific code changes require examples of such changes as training data~\cite{oopsla2019}.
Finally, researchers systematically study when and how developers perform specific kinds of changes to increase our understanding of development practices~\cite{negara2014mining,Rak-amnouykit2020,Nguyen2019,ase2020}.

Unfortunately, there currently is no efficient and effective technique for systematically searching large version histories for specific kinds of changes.
The solutions proposed in the above Stack Overflow post are all based on matching regular expressions against raw diffs.
However, searching for anything beyond the most simple change patterns with a regular expression is cumbersome and likely to result in irrelevant code changes.
Another existing technique is GitHub Search,\footnote{\url{https://github.com/search}} which allows for searching through commits using free-form queries that are matched, e.g., against commit messages.
However, both regular expressions and GitHub Search have significant drawbacks when searching for specific code changes, as we show in a user study.
Finally, previous research proposes techniques that linearly scan version histories for specific patterns~\cite{Fluri2007,kawrykow2011non,pan2009toward,Lawall2016prequel}.
However, due to their linear design, these techniques do not scale well to searching through hundreds of thousands of changes in a short time.

This paper presents \name{}, a scalable and precise search engine for code changes.
\name{} is enabled by three key contributions.
First, we design a query language that is intuitive to use and easy to adapt to different programming languages.
The query language extends the target programming language with wildcards and placeholders that abstract specific syntactic categories, e.g., expressions.
Second, to ensure scalability, the approach is split into an indexing part, which maps code changes into a feature space, and a retrieval part, which matches a given query in the feature space. We design specific features for code changes, extracting useful information to match different changes on source code. 
Finally, to ensure precision, i.e., that a found code change indeed fits the given query, a crucial part of the approach is to match candidate code changes against the given query.
We present an efficient algorithm that checks if a query can be expanded into a code change.

Our approach supports the different usage scenarios we envision \name{} to be useful for.
First, the approach supports users interested in finding \emph{one} specific code change, e.g., when searching through the history of their own project to find some change done by a colleague.
In this scenario, similar to a classical web search engine, the user will consider only the first few search results and stop inspecting them as soon as the expected code change is found.
Second, \name{} supports users interested in finding \emph{multiple} code changes, e.g., when searching through a set of popular open-source projects to find examples of typical ways to refactor a specific API usage.
In this scenario, the user will inspect the ranked list of search results until having seen a sufficient number of examples.
Third, the approach supports users interested in finding \emph{many} code changes, e.g., to build a large-scale dataset to train a neural model.
In this scenario, the user can formulate and fine-tune the query through the interactive user interface of \name{}, and then download all matching results at once into a file.
Finally, \name{} can also be configured to retrieve \emph{all} code changes that match a query, e.g., to quantify how often specific changes occur in practice.
In this scenario, the user turns off the indexing and retrieval part of the approach, and instead runs the precise matching of a query against all code changes.\footnote{As shown in the evaluation, guaranteeing to find \emph{all} matching code changes comes at the cost of efficiency, as it requires a linear search through all code changes in the corpus.}

\name{} is designed in a mostly language-agnostic way, making it possible to apply the approach to different languages.
In particular, we restrict ourselves to a very lightweight static analysis of code changes.
The query language and parts of the search algorithm build upon the context-free grammar of the target programming language.
As a proof-of-concept, \name{} currently supports three widely used languages: Java, JavaScript, and Python.

Our approach relates to work on searching for code, which retrieves code snippets that match keywords~\cite{sourcerer,Gu2018}, test cases~\cite{reissCodeSearch}, or partial code snippets~\cite{Luan2019,kim2018facoy}.
While code search engines often have a design similar to ours, i.e., based on indexing and retrieval, they consider only a single snapshot of code, but not code changes.
Other related work synthesizes an edit program from one or more code changes~\cite{Fluri2007,Falleri2014,Rolim2017,Gao2020,Erdweg2021} and infers recurring code change patterns~\cite{DBLP:conf/pldi/PaletovTRV18,Nguyen2019}.
Starting from concrete changes, these approaches yield abstractions of them.
Our work addresses the inverse problem: given a query that describes a set of code changes, find concrete examples that match the query.
Finally, our work relates to clone detection~\cite{kamiya2002ccfinder,Li2006,jiang2007deckard,roy2008nicad,sajnani2016sourcerercc}, as \name{} searches for code changes that resemble a query.
Our work differs from clone detection by considering code changes (and not individual snippets of code), by focusing on guaranteed matches instead of similar code, and by responding to queries quickly enough for interactive use.

We evaluate the effectiveness and scalability of \name{} with one million code changes in each of Java, Python, and JavaScript.
We find that the approach responds to queries within a few seconds, scaling well to large sets of code changes.
The search has a mean recall of 80.7\% for Java, 89.6\% for Python, and 90.4\% for JavaScript, which can be increased even further in exchange for a slight increase in response time. 
A user study shows that \name{} enables users to effectively retrieve code changes, clearly outperforming a regular expression-based search through raw diffs and GitHub Search.
As a case study to show the usefulness of \name{} for researchers, we apply the approach to gather a dataset of 74,903 bug fixes.

In summary, this paper contributes the following:
\begin{itemize}
\item A \emph{query language} that extends the target programming language with placeholders and wildcards, making it easy to adapt the approach to different languages.
\item A technique for searching for code changes that ensures \emph{scalability} through approximate, indexing-based retrieval, and that ensures \emph{precision} via exact matching.
\item Empirical evidence that the approach effectively finds thousands of relevant code changes, scales well to more than a million changes from different projects, and successfully helps users answer a diverse set of queries.
\end{itemize}

\smallskip
\noindent
The implementation and a web interface of DiffSearch are publicly available:
\begin{center}
\url{http://diffsearch.software-lab.org}
\end{center}

%For double-blind review, the site has been anonymized and does not perform any server-side tracking. The implementation will be available as open-source, and for now is available to reviewers\footnote{\url{http://u.pc.cd/9ItctalK}}, along with our experimental results.
\section{Example and Overview}\label{sec:overview}
\subsection{Motivating Example}\label{sec:motivation}

To illustrate the problem and how \name{} addresses it, consider the following example query.
The query searches for code changes that swap the arguments passed to a call that is immediately used in a conditional.
Such a query could be used to find fixes of swapped argument bugs~\cite{oopsla2017}.
\vspace{.99em}

{
\renewcommand{\arraystretch}{0.85}
\begin{tabular}{rlc}
&
\begin{minipage}[t]{13.5em}

\begin{Verbatim}[fontsize=\small]
if(ID<1>(EXPR<1>, EXPR<2>)){
  <...>

\end{Verbatim}
\end{minipage}\\
$\rightarrow$ &
\begin{minipage}[t]{15em}
\begin{Verbatim}[fontsize=\small]
if(ID<1>(EXPR<2>, EXPR<1>)){
  <...>
\end{Verbatim}
\end{minipage}
\end{tabular}
}
\vspace{.99em}

Our query language is an extension of the target programming language, Java in the example, and adds placeholders for some syntactic categories.
For example, the \code{ID<1>} placeholder matches any identifier, and the \code{EXPR<1>} placeholder matches any expression.
Instead of such placeholders, queries can also include concrete identifiers and literals, e.g., to search for specific API changes.

As the set of code changes to search through, suppose we have the following three examples, of which only the second matches the query:

\vspace{.3em}
\noindent \emph{Code change 1:}\\
%\begin{Verbatim}[fontsize=\small]
%if (check(a - 1, b)){ --> if (check(a - 1, c)) {
%\end{Verbatim}
\hspace*{-2.2em}
\begin{tabular}{rlcr}
	&
\begin{minipage}[t]{9.5em}
	\begin{Verbatim}[fontsize=\small]
if(check(a - 1, b)){
	\end{Verbatim}
\end{minipage}
&
\hspace{1em}$\rightarrow$
&
\begin{minipage}[t]{3em}
	\begin{Verbatim}[fontsize=\small]
if(check(a - 1, c)){
	\end{Verbatim}
\end{minipage}
\end{tabular}
\vspace{.3em}

\vspace{.3em}
\noindent \emph{Code change 2:}\\
%\begin{Verbatim}[fontsize=\small]
%if (isValid(x, y)) { --> if (isValid(y, x)) {
%\end{Verbatim}
\hspace*{-2.2em}
\begin{tabular}{rlcr}
	&
	\begin{minipage}[t]{11.1em}
		\begin{Verbatim}[fontsize=\small]
if(isValidPoint(x, y)){
			\end{Verbatim}
		\end{minipage}
		&
	\hspace{.7em}$\rightarrow$\hspace{-.8em}
		&
		\begin{minipage}[t]{2em}
			\begin{Verbatim}[fontsize=\small]
if(isValidPoint(y, x)){
				\end{Verbatim}
			\end{minipage}
		\end{tabular}
\vspace{.3em}

\vspace{.3em}
\noindent \emph{Code change 3:}\\
\hspace*{-2.2em}
\begin{tabular}{rlcr}
	&
	\begin{minipage}[t]{10.5em}
		\begin{Verbatim}[fontsize=\small]
while(var > k - 1){
  sum += count(var);
			\end{Verbatim}
		\end{minipage}
		&
		$\rightarrow$\hspace{-.7em}
		&
		\begin{minipage}[t]{6em}
			\begin{Verbatim}[fontsize=\small]
while(var > k){
  sum += 2 * count(var);
				\end{Verbatim}
			\end{minipage}
		\end{tabular}

\subsection{Problem Statement}

An important design decision is the granularity of code changes to consider.
The options range from changes of individual lines, which would limit the approach to very simple code changes, to entire commits, which may span multiple files, several dozens of lines~\cite{DBLP:conf/iwpc/AlaliKM08}, often containing multiple entangled logical changes~\cite{kawrykow2011non,DBLP:conf/icse/BarnettBBL15,DBLP:journals/ese/HerzigJZ16,Partachi2020a}.
We opt for a middle ground between these two extremes and consider code changes at the level of ``hunks'', i.e., consecutive lines that are added, modified, or removed together.
%This level of granularity may include multiple statements that are changed together, while capturing likely related code locations.

\begin{definition}[Code change]
\label{def:code change}
A code change $c \rightarrow c'$ consists of two pieces of code, each of which is a sequence $[l_1,..,l_m]$ of consecutive lines of code extracted from a file in the target language.
\end{definition}

%In principle, a piece of code must be syntactically complete.
%In practice, our implementation also handles many syntactically incomplete pieces of code, e.g., lines that open a block without closing it (\Cref{sec:trees}), as illustrated in the motivating example.

\begin{definition}[Query]
\label{def:query}
A query $q \rightarrow q'$ consists of two patterns, which each are a sequence $[l_1,..,l_m]$ of lines of code in an extension of the target programming language.
The language extension adds wildcards, a special ``empty'' symbol, and placeholders for specific syntactic categories, e.g., to match an arbitrary expression or identifier.
\end{definition}

Given these two ingredients, the problem we address is:
\begin{definition}[Search for code changes]
\label{def:search}
Given a set $C$ of code changes and a query $q \rightarrow q'$, find a set $M \subseteq C$ of code changes such that each $(c \rightarrow c') \in M$ matches $q \rightarrow q'$.
We say that a code change $c \rightarrow c'$ matches a query $q \rightarrow q'$ if there exists an expansion of the placeholders and wildcards in $q \rightarrow q'$ that leads to $c \rightarrow c'$.
\end{definition}

By ensuring that, for any retrieved code change, the query can be expanded to the code change, \name{} guarantees that every result of a search precisely matches the query.

\subsection{Main Idea of the Approach}
\label{sec:overview overview}

\begin{figure}
	\centering
	\includegraphics[width=0.98\linewidth]{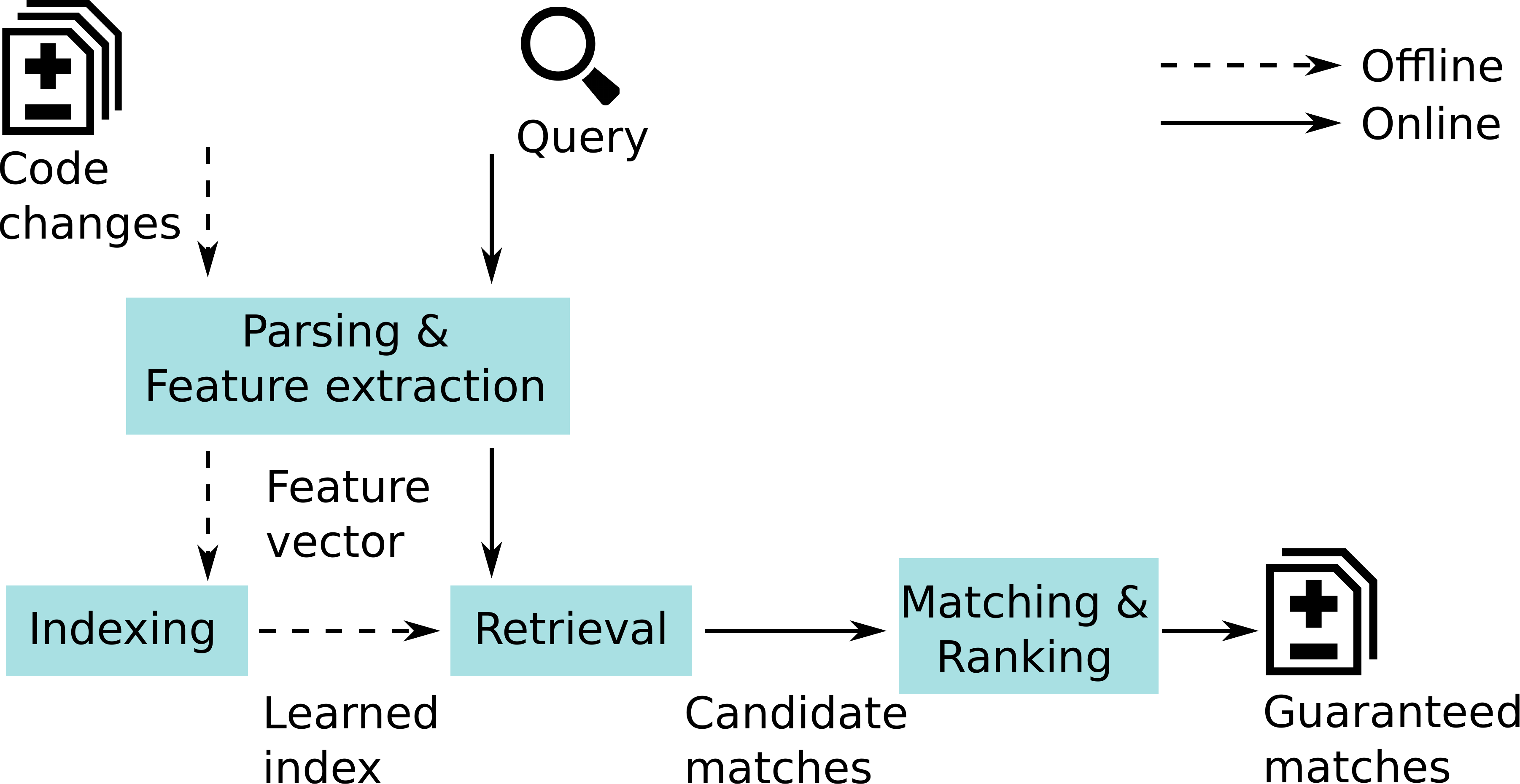}
	\caption{Overview of the approach.}
	\label{fig:overview}
\end{figure}

\name{} consists of four components that are used in an offline and an online phase as illustrated in \Cref{fig:overview}.
In the offline phase, the approach analyzes and indexes a large set of code changes.
The \emph{Parsing \& Feature extraction} component of the approach parses and abstracts concrete code changes and queries into a set of features, mapping both into a common feature space.
For our example query in Section~\ref{sec:motivation}, the features encode, e.g., that a call expression appearing within the condition of an if statement is changed and that the changed call has two arguments.
To enable quickly searching through hundreds of thousands of code changes, the \emph{Indexing} component of \name{} indexes the given feature vectors~\cite{johnson2019billion} once before accepting queries.

In the online phase, the input is a query that describes the kind of code changes to find.
Based on the pre-computed index and the feature vector of a given query, the \emph{Retrieval} component retrieves those code changes that are most similar to the query.
For our motivating example, this yields Code change~1 and Code change~2 because both change the arguments passed to a call.
The similarity-based retrieval does not guarantee precision, i.e., that each candidate code change indeed matches the query.
The \emph{Matching \& Ranking} component of \name{} removes any candidates that do not match the query by checking whether the placeholders and wildcards in the query can be expanded into concrete code in a way that yields the candidate code change.
For our example, matching will eliminate Code change~1, as it does not swap arguments, and eventually returns Code change~2 as a search result to the user.
 
\section{Approach}

This section presents the approach in detail.
Before going through the four components introduced in \Cref{sec:overview overview}, we define the query language to specify what kind of code changes to search for.

\subsection{Query Language}
\label{sec:queryingLanguage}

To search for specific kinds of code changes, \name{} accepts queries that describe the code before and after the change.
Our goal is to provide a query language that developers can learn with minimal effort and that supports all constructs of the target programming language.
We initially considered three possible kinds of code search queries, as classified by Di Grazia et al.~\cite{DiGrazia2022survey}.
First, natural language queries, which are easy to type but inherently imprecise.
Second, programming language queries, which require knowing the programming language and are precise.
Third, custom languages that are often the most precise, but they may impose some effort to learn the new language~\cite{DiGrazia2022survey}.

Comparing the different options and considering the envisioned users of our approach, we design the query language of \name{} as an extension of the target programming language.
That is, the query language includes all rules of the target programming language and additional features useful for queries.
As our approach can support different target languages, this means that there is a different query language for each target language, each extending the target language with search-related keywords.
That is, a user who is already familiar with the target programming language needs to learn only a handful of new keywords for using \name{}.

\renewcommand{\syntleft}{\normalfont\itshape}
\renewcommand{\syntright}{}

\begin{figure}
    \setlength{\grammarparsep}{.2em}
    \setlength{\grammarindent}{8em}
    \small
	\begin{grammar}
		<Query> ::= <Snippet> $\rightarrow$ <Snippet>

		<Snippet> ::= <Stmt>*  | <Expression>
		         | _

		<Stmt> ::= $\langle$...$\rangle$
		      | (Target language rules)

		<Expression> ::= EXPR
		            | EXPR$\langle$<Number>$\rangle$
		            | $\langle$...$\rangle$ | (Target language rules)

		<AssignOperator> ::= OP
		          |  OP$\langle$<Number>$\rangle$
		          | (Target language rules)

		<BinaryOperator> ::= binOP
			          | binOP$\langle$<Number>$\rangle$
			          | (Target language rules)

        <UnaryOperator> ::= unOP
        | unOP$\langle$<Number>$\rangle$
        | (Target language rules)

		<Identifier> ::= ID
		          | ID$\langle$<Number>$\rangle$
		          | (Target language rules)

		<Literal> ::= LT
		         | LT$\langle$<Number>$\rangle$
		         | (Target language rules)
	\end{grammar}
	\caption{Simplified grammar of queries. Non-terminals are in \textit{italics}.}
	\label{fig:grammar}
\end{figure}

\Cref{fig:grammar} shows the grammar of our query language.
A query consists of two sequences of statements, which describe the old and new code, respectively.
The syntax for statements is inherited from the target programming language and not shown in the grammar.
Instead of a regular code snippet, a query may contain an underscore to indicate the absence of any code, which is useful to describe code changes that insert or remove code.
The grammar extends the target language by adding placeholders for specific syntactic entities, namely expressions, operators, identifiers, and literals.
For each such entity, a query can either describe with an unnamed placeholder that there should be any such entity, e.g., \code{EXPR} for any expression, or repeatedly refer to a specific entity with a named placeholder, e.g., using \code{EXPR<1>} and \code{EXPR<2>}.
Named placeholders will be bound to the same entity across the entire query, e.g., to say that the same expression \code{EXPR<1>} must appear on both sides. We also introduce the wildcard \code{<...>} that matches any statement, any expression, or nothing at all.
%% MP: Not true anymore:
%For queries that contain multiple named placeholders, \name{} ensures that each named placeholder of the same kind, e.g., \code{EXPR<1>} and \code{EXPR<2>}, is bound to a different entity.
%That way, a query can express that specific part of the code must change.

\begin{table}
	\caption{Examples of Java changes and matching queries.}
	\label{tab:grammarExamples}
	\setlength{\tabcolsep}{5pt}
	\small
	\begin{tabular}{@{}ll@{}}
		\toprule
		Code change&DiffSearch query \\
		\midrule
		\begin{minipage}{8em}
		\small
		\begin{Verbatim}
- evt.trig();
		\end{Verbatim}
		\end{minipage}
		&
		\begin{minipage}[t]{8em}
		\small
		\begin{Verbatim}
ID.ID();
		\end{Verbatim}
		\end{minipage}
		\hspace{-.3em}$\rightarrow$\hspace{.3em}
	\begin{minipage}[t]{8em}
	\begin{Verbatim}
_
	\end{Verbatim}
	\end{minipage}
		\\
		\midrule
		\begin{minipage}[t]{6.5em}
		\small
		\begin{Verbatim}
- if (x > 0)
-   y = 1;
+ if (x < 0)
+   y = 0;
		\end{Verbatim}
		\end{minipage}
		&
		\begin{minipage}[t]{8em}
		\small
		\begin{Verbatim}
if (EXPR)
  ID OP LT;
		\end{Verbatim}
		\end{minipage}
		\hspace{-.3em}$\rightarrow$\hspace{.3em}
	\begin{minipage}[t]{8em}
	\small
	\begin{Verbatim}
if (EXPR)
  ID OP LT;
	\end{Verbatim}
	\end{minipage}
		\\
		\midrule
		\begin{minipage}[t]{6.5em}
		\small
		\begin{Verbatim}
- run(k);
- now(k);
+ runNow(k);
		\end{Verbatim}
		\end{minipage}
		&
		\begin{minipage}[t]{8em}
		\small
		\begin{Verbatim}
run(EXPR<0>);
now(EXPR<0>);
		\end{Verbatim}
		\end{minipage}
		\hspace{-.3em}$\rightarrow$\hspace{.3em}
	\begin{minipage}[t]{9em}
	\small
	\begin{Verbatim}
runNow(EXPR<0>);
	\end{Verbatim}
	\end{minipage}
		\\
		\bottomrule
	\end{tabular}
\end{table}

To illustrate the query language, \Cref{tab:grammarExamples} gives a few examples of code changes and a corresponding query that matches the code change.
The first two examples use unnamed placeholders, e.g., to match arbitrary identifiers.
The third example uses a named placeholder:
The \code{EXPR<0>} in both the old and new part of the query means that this expression, here \code{k}, remains the same despite the code change, which replaces two calls with one.

\subsection{Tree-based Representation of Code Changes and Queries}
\label{sec:trees}

One goal of \name{} is to be mostly language-agnostic, making it possible to apply the approach to different programming languages.
Our current version supports Java, JavaScript, and Python.
To this end, the approach represents code changes and queries using a parse tree, i.e., a representation that is straightforward to obtain for any programming language.
The benefit of parse trees is that they abstract away some details, such as irrelevant whitespace, yet provide an accurate representation of code changes.

\begin{figure*}
	\centering
	\includegraphics[width=.9\linewidth]{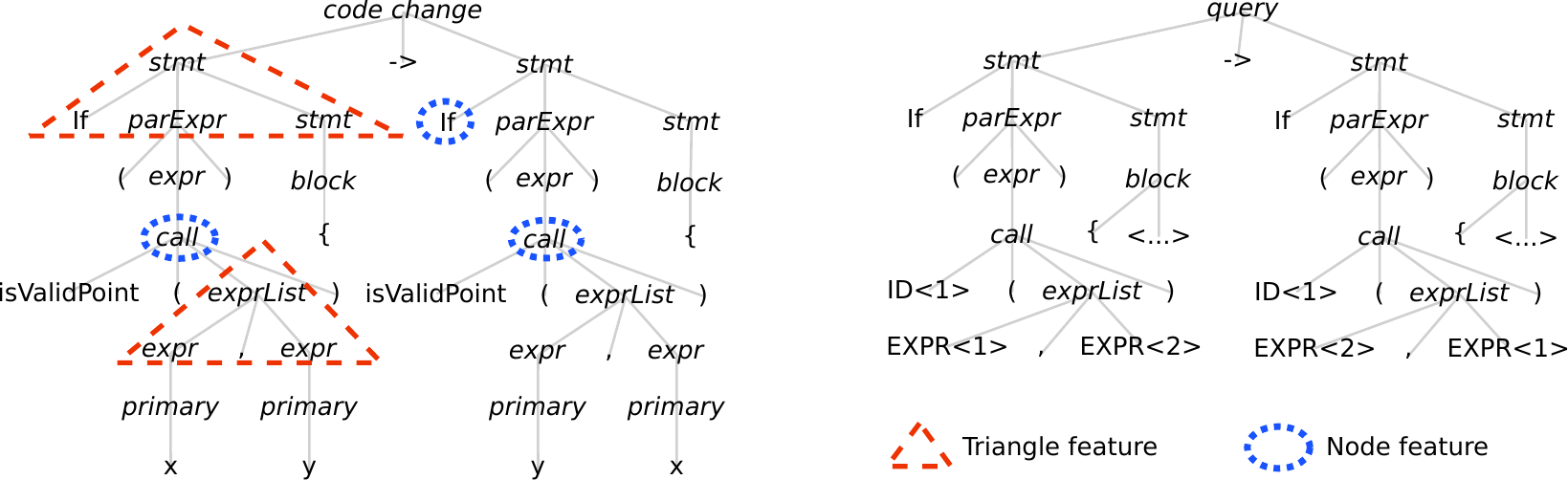}
	\caption{Parse tree representations of Code change~2 (left) and the query from \Cref{sec:overview} (right). Only some of all considered features are highlighted for illustration.}
	\label{fig:trees}
\end{figure*}

To represent a set of commits in a version history as pairs of trees, \name{} first splits each commit into hunks, which results in a set of code changes (\Cref{def:code change}).
The approach then parses the old and new code of a hunk using the programming language grammar into a single tree that represents the code change.
Likewise, to represent a query, \name{} parses the query into a parse tree using our extension of the grammar (\Cref{fig:grammar}).
For example, \Cref{fig:trees} shows the parse trees of a change and a query.
The change on the left corresponds to Code change~2 from \Cref{sec:overview}, which swaps \code{x} and \code{y} of a call to \code{isValidPoint}.
Note that code edits that do not cause any change of the parse tree, e.g., because only semantically irrelevant whitespace gets changed, are not considered as code changes and ignored by \name{}.

An interesting challenge in parsing code changes and queries is syntactically incomplete code snippets.
For example, the code changes in \Cref{sec:overview} open a block with \code{\{} but do not close it with \code{\}}, because the line with the closing curly brace was not changed.
\name{} addresses this challenge by relaxing the grammar of the target language so that it accepts individual code lines even when they are syntactically incomplete.
For example, we relax the grammar to allow for unmatched parentheses and partial expressions.
%Revising a grammar to support syntactically incomplete snippets is a one-time effort that we found to be easy to perform.

As a potential alternative to parse trees, we considered and eventually decided against abstract syntax trees (ASTs).
While ASTs are a suitable representation, e.g., for compilers, they abstract away too many syntactic details that may be relevant in \name{}.
For example, consider the following code change that adds parentheses to make a complex expression easier to read:
%\begin{Verbatim}[fontsize=\small]
%- flag = stop || root && live;
%+ flag = stop || (root && live);
%\end{Verbatim}

\begin{tabular}{ll}
    &
	\begin{minipage}[t]{10em}
		\begin{Verbatim}[fontsize=\small]
flag = alive || x && y;
		\end{Verbatim}
	\end{minipage}
	\\
	$\rightarrow$
	&
	\begin{minipage}[t]{6em}
		\begin{Verbatim}[fontsize=\small]
flag = alive || (x && y);
		\end{Verbatim}
	\end{minipage}
\end{tabular}
\vspace{.3em}

Because the added parentheses preserve the semantics of the expression, they are abstracted away in a typical AST, i.e., the old and new code have the same AST.
As a result, an AST-based representation could neither represent this change nor a query to search for it.

\subsection{Extracting Features}
\label{sec:features}

Based on the tree representation of code changes and queries, the feature extraction component of \name{} represents each tree as a set of features.
The goal of this step is to enable quickly searching through hundreds of thousands of code changes.
By projecting both code changes and queries into the same feature space, we enable the approach to compare them efficiently.
An alternative would be to pairwise compare each code change with a given query~\cite{Fluri2007,Lawall2016prequel}.
However, such a pairwise comparison would require an amount of computation time that is linear w.r.t.\ the number of code changes, which would negatively affect the efficiency of searching through many code changes.

\name{} uses two kinds of features.
The first kind of feature is \emph{node features}, which encode the presence of a node in the parse tree.
For the example in \Cref{fig:trees}, the dotted, blue lines show three of the extracted node features.
The second kind of feature is \emph{parse tree triangles}, which encode the presence of a specific subtree.
Each parse tree triangle is a tree that consists of a node and all its descendants up to some configurable depth.
We use a depth of one as a default, i.e., a triangle contains a node and its immediate child nodes.
For the example in \Cref{fig:trees}, the dashed, red lines highlight two of the extracted triangles.
The triangle at the top encodes the fact that there is an if statement, while the other triangle encodes the fact that the code contains an expression list with exactly two expressions.
The two kinds of features complement each other because node features encode information about individual nodes, including identifiers and operators, whereas parse tree triangles represent how nodes are connected.

For each code change or query, the approach extracts a separate set of features for the old and the new code.
With this separation, the features encode whether specific code elements are added or removed in a code change.
The feature sets for code changes and queries are constructed in the same way, except that \name{} removes node features for placeholder nodes, e.g., \code{ID} or \code{EXPR}, from the query.
The rationale is that we want the features of a query to be a subset of the features of a matching code change, but placeholder nodes never appear in code changes.

Different code changes and queries yield different numbers of features.
To efficiently compare a given query against arbitrary code changes, \name{} represents all features of a code change or query as a fixed-size feature vector.
The feature vector is a binary vector of length $l_{\mathit{n}} + l'_{\mathit{n}} + l_{\mathit{tri}} + l'_{\mathit{tri}}=l$, where $l_{\mathit{n}}$ and $l'_{\mathit{n}}$ are the number of bits to represent the node features of the old and new code, respectively, and likewise for $l_{\mathit{tri}}$ and $l'_{\mathit{tri}}$ for the parse tree triangle features.
We use $l=$ \emph{1,000} by default, dividing it equally among the four components, which strikes a balance between representing a diverse set of features and efficiency during indexing and retrieval.
Section~\ref{sec:eval parameters} evaluates different sizes for the feature vector length. 

\begin{algorithm}[tb]
%	\small
	\begin{algorithmic}[1]
		\Require{Set $F$ of features, target size $l_{\mathit{target}}$}
		\Ensure{Feature vector $v$}
		\State $v \leftarrow$ vector of $l_{\mathit{target}}$ zeros
		\ForAll{$f \in F$}
			\State $h \leftarrow \mathit{hash}(f)$
			\State $v[h \bmod{} l_{\mathit{target}}] \leftarrow 1$\label{line:mod1}
		\EndFor
		\State \textbf{return} $v$
	\end{algorithmic}
	\caption{Represent features as fixed-size vector.}\label{alg:featureVector}
\end{algorithm}

\Cref{alg:featureVector} summarizes how \name{} maps a set $F$ of features into a fixed-size vector $v$.
The algorithm computes a hash function over the string representations of individual nodes in a feature, sums up the hash values into a value $h$, and sets the $h$-th index of the feature vector to one.
To ensure that the index is within the bounds of $v$, line~\ref{line:mod1} performs a modulo operation.
For each code change or query, the algorithm is invoked four times to map each of the four feature sets into a fixed-size vector: parent-child and triangle features, for both the old and new code.

%\begin{figure}[H]
%	\centering
%	\includegraphics[scale=0.5]{images/featureVector}
%	\caption{An abstraction of the feature vectors.}
%	\label{fig:featureVector}
%\end{figure}

\subsection{Indexing and Retrieving Code Changes}
\label{sec:indexingRetrieval}

To prepare for responding to queries, \name{} runs an offline phase that indexes the given set of code changes.
The indexing and retrieval components of the approach build on FAISS, which is prior work on efficiently searching for similar vectors across a large set of vectors~\cite{johnson2019billion}.
In the first step of the offline phase, \name{} parses all code changes and stores the parse trees on disk.
In the second step, \name{} generates the feature vectors of the code changes using the corresponding parse trees.
Given the set $V_{\mathit{changes}}$ of feature vectors of all code changes, the approach computes an index into these vectors.

After the offline indexing phase, \name{} accepts queries.
For a given query, the approach computes a feature vector $v_{\mathit{query}}$ (\Cref{sec:features}), and then uses the index to efficiently retrieve the most similar feature vectors of code changes.
FAISS allows for efficiently answering approximate nearest neighbor queries, without comparing the query against each vector in $V_{\mathit{changes}}$.
The nearest neighbors are based on the L2 (Euclidean) distance.
To ensure that the presence of matching features is weighted higher than the absence of features, we multiply $v_{\mathit{query}}$ by a constant factor $\frac{l}{2}+1$ before running the nearest neighbor query.
To illustrate this decision consider an example with three feature vectors: A query $v_Q=(0,0,1)$, a potential match $v_P=(1,1,1)$ with the third feature in common, and a mismatch $v_M=(0,0,0)$.
Naively computing the Euclidean distances yields $d(v_Q,v_P)= \sqrt{2}$ and $d(v_Q,v_M) = \sqrt{1}$, i.e., the mismatch would be closer to the query than the potential match.
To avoid this scenario, the query vector should be $v_Q=(0,0,m)$ such that $d(v_Q, v_P) < d(v_q, v_M)$.
Solving this inequality gives $m > \frac{l}{2}$, which we achieve by multiplying the original $v_Q$ with $\frac{l}{2}+1$.  
For the example, after multiplying $v_Q$ with the constant factor $\frac{3}{2}+1$, we have $d(v_Q,v_P)= \sqrt{4.25}$ and $d(v_Q,v_M) = \sqrt{6.25}$, i.e., the potential match is now closer to the query than to the mismatch.

The approach retrieves the $k$ most similar code changes for a given query.
Setting the value of $k$ allows users to control the trade-off between efficiency and recall.
For example, if a user is interested in finding as many code changes as possible, a larger $k$ should be used.
In the extreme case, \name{} can also be used without the feature-based retrieval (equivalent to $k=\infty$), which will reduce the approach to linearly searching through all code changes, but guarantees to find each matching code change.
We use $k=$~\emph{5,000} by default, and Section~\ref{sec:eval parameters} evaluates other values.
The retrieved candidate code changes are ranked based on their L2 distance to the query, computed by FAISS, and we use this ranking to sort the final search results shown to a user.

\subsection{Matching of Candidate Search Results}
\label{sec:pruning}
Given the $k$ candidate code changes retrieved for a given query as described in \Cref{sec:indexingRetrieval}, \name{} could return all of them to the user.
However, the feature-based search does not guarantee precision, i.e., that all the retrieved code changes indeed match the query.
One reason is that the features capture only local information, but do not encode the entire parse tree in a lossless way.
Another reason is that the features do not encode the semantics of named placeholders, i.e., they cannot ensure that placeholders are expanded consistently across the old and new code.

To guarantee that all code changes returned in response to a query precisely match the query, the matching component of \name{} takes the candidate search results obtained via the feature-based retrieval and checks for each candidate whether it indeed matches the query.
Intuitively, a code change matches a query if the placeholders and wildcards in the query can be expanded in a way that yields code identical to the code change or some subset of the code change.
More formally, we define this idea as follows:
\begin{definition}[Match]
\label{def:match}
Given a code change $c \rightarrow c'$ and a query $q \rightarrow q'$, let $t_c, t_{c'}, t_q, t_{q'}$ be the corresponding parse trees.
The code change matches the query if
\begin{itemize}
\item $t_q$ can be expanded into some subtree of $t_c$ and
\item $t_{q'}$ can be expanded into some subtree of $t_{c'}$
\end{itemize}
so that all of the following conditions hold:
\begin{itemize}
\item Each placeholder is expanded into a subtree of the corresponding syntactic entity.
\item All occurrences of a named placeholder are consistently mapped to identical subtrees.
\item Each wildcard is expanded to an arbitrary, possibly empty subtree.
\end{itemize}
\end{definition}

For example, consider the query and code change in \Cref{fig:trees} again.
They match because the tree on the right can be expanded into the tree on the left.
The expansion maps the named placeholders \code{ID<1>} to \code{isValidPoint}, \code{EXPR<1>} to the subtree that represents \code{x}, and \code{EXPR<2>} to the subtree that represents \code{y}.
Moreover the wildcards in the query are both mapped to the empty tree.
As an example of a code change that does not match this query, consider Code change~1 from \Cref{sec:overview} again.
The parse tree of the query cannot be expanded into the parse tree of that code change because there is no way of expanding the query tree while consistently mapping \code{EXPR<1>} and \code{EXPR<2>} to the three method arguments \code{a-1}, \code{b}, and \code{c}.

%The reason for saying ``some subtree of $t_c$'' in \Cref{def:match} is that \name{} should also retrieve code changes that, in addition to code as described in the query, contains other code around it.
%This way, the approach can find relevant changes that are surrounded by other, irrelevant changes.
%%
%For example, the query in \Cref{fig:trees} would find Code change~2 even if another statement is changed just before the if statement.

To check whether a candidate code change indeed matches the given query, \name{} compares the parse tree of the query with the parse tree of the code change in a top-down, left-to-right manner.
The basic idea is to search for a mapping of nodes in the query tree to nodes in the parse tree that consistently maps named placeholders to identical subtrees.
On top of this basic idea, the matching algorithm faces two interesting challenges.
We illustrate the challenges with the following query, which searches for code changes where two call statements get replaced by an assignment of a literal to an identifier. The following example shows the query on the left and a matching code change on the right:

%\vspace{.5em}
%\begin{tabular}{@{}lcr@{}}
%\begin{minipage}{3em}
%\begin{Verbatim}[fontsize=\small]
%ID();
%<...>
%ID();
%\end{Verbatim}
%\end{minipage}
%&
%$\rightarrow$
%&
%\begin{minipage}{6em}
%\begin{Verbatim}[fontsize=\small]
%ID = LT;
%\end{Verbatim}
%\end{minipage}
%\end{tabular}
%\vspace{.5em}\\
%Consider the following code change, which matches the query:
%\begin{Verbatim}[fontsize=\small]
%- foo();
%- bar();
%- baz();
%+ x = 5;
%+ foo();
%+ y = 7;
%\end{Verbatim}

%\vspace{.5em}
%\begin{tabular}{@{}lcr@{}}
%	\begin{minipage}{3em}
%		\begin{Verbatim}[fontsize=\small]
%foo();
%bar();
%baz();
%		\end{Verbatim}
%	\end{minipage}
%	&
%	$\rightarrow$
%	&
%	\begin{minipage}{6em}
%		\begin{Verbatim}[fontsize=\small]
%x = 5;
%foo();
%y = 7;
%		\end{Verbatim}
%	\end{minipage}
%\end{tabular}
%\vspace{.5em}\\
\vspace{.3em}
\begin{tabular}{@{}lcr@{}}
	\begin{minipage}{3em}
		\begin{Verbatim}[fontsize=\small]
ID();
<...>
ID();
		\end{Verbatim}
	\end{minipage}
	&
	$\rightarrow$
	&
	\begin{minipage}{6em}
		\begin{Verbatim}[fontsize=\small]
ID = LT;
		\end{Verbatim}
	\end{minipage}
	\hspace{2em}
	\begin{minipage}{3em}
	\begin{Verbatim}[fontsize=\small]
foo();
bar();
baz();
	\end{Verbatim}
\end{minipage}

$~\rightarrow$

\begin{minipage}{6em}
	\begin{Verbatim}[fontsize=\small]
 x = 5;
 foo();
 y = 7;
	\end{Verbatim}
\end{minipage}
\end{tabular}
\vspace{.3em}

The first challenge is because queries are allowed to match parts of a change, which is useful to find relevant changes surrounded by other, irrelevant changed code.
While useful, this property of queries also implies that the query may match at multiple places within a given code change.
In the above example, the \code{ID = LT;} part of the query may match both \code{x = 5;} and \code{y = 7;}.
The second challenge is because queries may contain wildcards (\code{<...>}), which is useful to leave parts of a query unspecified.
Wildcards can match none, one, or multiple statements or expressions, and hence, they may cause a single query to match in multiple ways.
For the above example, the wildcard could be between the calls of \code{foo} and \code{baz}, between the calls of \code{foo} and \code{bar}, or between the calls of \code{bar} and \code{baz}.
Because of these two challenges, matching must consider different ways of mapping a query onto a code change, which results in a search space of possible matches that must be explored.

\begin{algorithm}[tb]
	\caption{Check if a code change matches a query.}
	\label{alg:match}
	\small
	\begin{algorithmic}[1]
		\Require{Code change $c \rightarrow c'$ and query $q \rightarrow q'$}
		\Ensure{True if they match, False otherwise.}
		\State $t_c, t_{c'} \leftarrow \mathit{parse}(c \rightarrow c')$
		\State $t_q, t_{q'} \leftarrow \mathit{parse}(q \rightarrow q')$

		\State $N_{\mathit{toMatch}} \leftarrow (\mathit{allNodes}(q) \cup \mathit{allNodes}(q')) \setminus \mathit{wildcards}$
		\State $W \leftarrow \mathit{candidateMappings}(t_c, t_{c'}, t_q, t_{q'})$

		\While{$W$ is not empty}
			\State $M \leftarrow$ Take a mapping from $W$
			\State $n_q \leftarrow \mathit{nextUnmatchedNode}(M, t_q, t_{q'})$

			\State $n_{pq} \leftarrow$ Parent of $n_q$
			\State $n_{pc} \leftarrow$ Look up $n_{pq}$ in $M$

			\For{$c$ \mbox{\textbf{in}} all not yet matched children of $n_{pc}$}
				\If{$\mathit{canAddToMap}(M, c, n_q)$}
					\State $M' \leftarrow$ Copy of $M$ with $n_q \mapsto c$
					\If{$\mathit{keys}(M') \cap N_{\mathit{toMatch}} = \emptyset$\\
					\hspace{5.1em} \mbox{\textbf{and}} $\mathit{isValid}(M, t_c, t_{c'}, t_q, t_{q'})$}
						\State \textbf{return} true \label{line:isMatch}
					\EndIf
				\Else
					\State Add $M'$ to $W$
				\EndIf
			\EndFor
		\EndWhile
	\end{algorithmic}
\end{algorithm}

\name{} addresses these challenges in \Cref{alg:match}, which checks whether a given query and code change match.
The algorithm starts by parsing the code change into trees $t_c$ and $t_{c'}$, which represent the old and new part of the change, and likewise for the query.
The core of the algorithm is a worklist-based search through possible mappings between nodes in the parse tree of the query and nodes in the parse tree of the code change.
These mappings are represented as a map $M$ from nodes in the query trees to nodes in the code change trees.
Each mapping $M$ in the worklist $W$ represents a possible way of matching the query against the code change.
To determine whether all nodes in the query have been successfully mapped, the algorithm maintains a set $N_{\mathit{toMatch}}$ of all the nodes in the query that must be matched.
The algorithm explores mappings in $W$ until it either finds a mapping that covers all nodes in $N_{\mathit{toMatch}}$, or until it has unsuccessfully explored all mappings in $W$.

\Cref{alg:match} relies on several helper functions.
One of them, $\mathit{candidateMappings}$, computes the starting points for the algorithm by returning all possible mappings of the roots of $t_q$ and $t_{q'}$ to nodes in the code change trees.
The $\mathit{nextUnmatchedNode}$ function performs a top-down, left-to-right pass through the query trees to find a node that is not yet in the current map $M$.
The $\mathit{canAddToMap}$ function checks if adding a mapping $n_q \mapsto c$ is consistent with an already existing map $M$.
Specifically, it checks that $n_q$ is not yet among the keys of $M$, that $c$ is not yet among the values of $M$, and that the two nodes are either identical non-placeholder nodes or that $n_q$ is a placeholder that can be consistently mapped to $c$ as specified in \Cref{def:match}.
Finally, the helper function $\mathit{isValid}$ checks whether a mapping $M$ that covers all to-be-matched nodes ignores nodes in the change tree only when there is a corresponding wildcard in the query tree. The algorithm postpones this check to $\mathit{isValid}$ to reduce the total number of mappings to explore.

%candidateMappings (c, c', p, p'):
%for p, find all nodes in c of same (e.g., expression vs expression) or compatible (e.g., EXPR ev expression) type
%same for p' and c'
%return all pairs of p->someNodeInC, p'->someNodeInC'
%
%nextUnmatchedNode(m, p, p'):
%top-down, left-to-right pass through query to find node not yet in m

%canAddToMap(m, c, n):
%checks that
% c not yet in keys of m,
% n not yet in values of m,
% if both normal nodes then same,
% if c is unnamed placeholder, n is matching entity
% if c is named placeholder, n is matching entity and if previously used, the mapped to identical subtree

%main algo assumes tt wildcard between all sibling nodes,
%isValid checks if omitted nodes between matching nodes correspond to wildcards
%benefit: reduces search space in practice; otherwise would have to match every wildcard to all prefixes of not yet matched nodes; if n children of already matched node not yet matched and wildcard in query, would need to consider 0, 1, ..., n of the following nodes

%isValid(m, c, c', p, p'):
%for both c,p and c',p':
%  top-down, left-to-right check tt p can be expanded into c while mapping wildcards in p to on or more subsequent sibling nodes in c

Matching a single code change against a query might cause the algorithm to explore many different mappings, and \name{} typically invokes \Cref{alg:match} not only once but for tens or hundreds of candidate search results.
To ensure that the approach responds to queries quickly enough for interactive usage, we optimize \Cref{alg:match} by pruning code changes that certainly cannot match a given query.
To this end, the approach checks if all leaf nodes in the parse tree of a query occur at least once in the parse tree of the code change.
For example, consider the following query, which searches for changes in the right-hand side of assignments to a variable \code{myVar}:\footnote{Because the \scode{myVar =}\hspace{.5em} part of the code remains the same, the query expresses that the literal captured by the unnamed placeholder \scode{LT} is changing.}

\vspace{.2em}
\hspace{2em}
\begin{tabular}{rlcr}
&
\begin{minipage}[t]{6em}
\begin{Verbatim}[fontsize=\small]
myVar = LT;
\end{Verbatim}
\end{minipage}
&
$\rightarrow$
&
\begin{minipage}[t]{6em}
\begin{Verbatim}[fontsize=\small]
myVar = LT;
\end{Verbatim}
\end{minipage}
\end{tabular}
\vspace{.3em}

If a code change does not include any token \code{myVar}, then the optimization immediately decides that the code change cannot match the query and skips \Cref{alg:match}, similar to Coccinelle~\cite{Lawall2018}.

%A reader may ask why not simply compare a query against all code changes using \Cref{alg:match}.
%The reason is that the computational complexity of that approach would be linear in the number of code changes, i.e., every query would be compared to hundreds of thousands of changes.
%Instead, \name{} retrieves the $k$ changes that are likely to be most relevant for a query and then checks only for this subset whether the code changes match the query.

\section{Implementation}
\label{sec:implementation}

We implement the \name{} idea in a practical search engine that supports multiple programming languages, currently Java, JavaScript, and Python.
To gather raw code changes, the implementation uses ''git log -p''.
For each change, a parse tree is created using ANTLR4,\footnote{https://www.antlr.org/} using the grammar of the target programming language, modified to support queries and to allow for syntactically incomplete code fragments (\Cref{sec:queryingLanguage}).
The indexing and retrieval components build on the FAISS library~\cite{johnson2019billion}, which supports efficient vector similarity queries for up to billions of vectors.
Once changes are indexed, the search engine is a server that responds to queries via one of two publicly available interfaces: a web interface for interactive usage and a web service for larger-scale usage, e.g., to create a dataset of changes.\footnote{\url{http://diffsearch.software-lab.org}}

\section{Evaluation}
%\todo{Restructure the section: 1) Recall, 2) Scalability, 3) User study, 4) Case study: bug fixing patterns, 5)sensitivity analysis}\\
Our evaluation focuses on six research questions:
\begin{itemize}
\item RQ1: What is the recall of \name{}? (Section~\ref{sec:eval recall})
%A high recall is important to ensure finding a sufficient number of relevant code changes.
%We measure recall with 80 queries for which we compute a ground truth of relevant code changes .
\item RQ2: How efficient and scalable is \name{}? (Section~\ref{sec:evalScalability})
%We address this question by indexing and searching through more than a million code changes.
\item RQ3: Does \name{} enable users to find relevant code changes more effectively than a regular expression-based search through raw diffs? (Section~\ref{sec:eval user study})
%We address this question in a user study that compares \name{} to a regular expression-based search through raw diffs.
\item RQ4: Is \name{} useful for finding examples of recurring bug fix patterns? (Section~\ref{sec:eval bug patterns})
%We address this question in a case study based on bug fix patterns formulated by prior work~\cite{Karampatsis2019a} (Section~\ref{sec:eval bug patterns}).
\item RQ5: How do parameters of the approach influence the results? (Section~\ref{sec:eval parameters})
%We address this question through a sensitivity analysis.
\item RQ6: How do queries and search results compare in terms of their size and absolute number? (Section~\ref{sec:eval query vs results})
\end{itemize}

For each of RQ1, RQ2, RQ5, and RQ6, we present results for all three currently supported target languages: Java, JavaScript, and Python.
For each language, we gather at least one million code changes from repositories that are among the top 100 of their language based on GitHub stars.
We compute the average size of the code change pair (old code and new code) in these datasets. The datasets do not contain commit messages, meta-information or code context, but only the removed and added lines, as represented in the diff. As a result, we count the number of '\textbackslash n' in each pair using the bash command "grep -o '\textbackslash n' dataset \(|\)  wc -l" and we find an average number of lines per each pair of 13.4, 8.2, and 7.3 for Java, Python and JavaScript, respectively. 
For RQ3 and RQ4, we focus on Java as the target language because RQ3 is based on a user study and because RQ4 builds on a Java dataset created by prior work~\cite{Karampatsis2019a}.
%We limit the length of the code changes to 500 characters.
The experiments are performed on a server with 48 Intel Xeon CPU cores clocked at 2.2GHz, 250GB of RAM, running Ubuntu~18.04.

\subsection{RQ1: Recall}
\label{sec:eval recall}
While the precision of \name{}'s results is guaranteed by design (Section~\ref{sec:pruning}),
the approach may miss code changes due to its feature-based search, which ensures scalability but may fail to include an expected code change into the candidate matches.
Additionally, \name{} only considers $k$ candidate changes, so it can find at most $k$ results even though queries could have more than $k$ matching code changes.

To establish a ground truth, we randomly sample code changes $c \rightarrow c'$ from all indexed Java, Python, and JavaScript code changes and formulate a corresponding query $q \rightarrow q'$ using the following four strategies. 
The \emph{as-is} strategy simply copies $c$ into $q$ and $c'$ into $q'$.
The \emph{less-placeholders} strategy replaces some of the identifiers, operators, and literals with corresponding placeholders or wildcards.
The \emph{more-placeholders} strategy, similarly, replaces the majority of the identifiers, operators, and literals.
Finally, the \emph{generalized} strategy replaces most or all of the identifiers, operators, and literals.
For each strategy and each programming language, we randomly sample 20 code changes and construct a query for each one.
We then compare each query against all 1,001,797 Java, 1,007,543 JavaScript, and 1,016,619 Python code changes using the matching component of \name{}.
While significantly slower than the feature-supported search that \name{} uses otherwise, this approach allows us to determine the set of all code changes expected to be found for a query, because Algorithm~\ref{alg:match} precisely computes whether a code change matches a query.
By design of \name{} (Section~\ref{sec:pruning}) and the way we construct the ground truth, the precision and the mean reciprocal rank (MRR) are 100\% and 1.0, respectively, and we hence do not report them in Table~\ref{tab:Recall}.

\begin{table}[]
	\centering
	\caption{Recall of \name{} across 80 queries per language.}
	\label{tab:Recall}
\setlength{\tabcolsep}{17pt}
\begin{tabular}{@{}lrrr@{}}
	\toprule
	Queries & Java & Python & JavaScript \\ \midrule
	As-is & 90.6\% & 100.0\% & 100.0\% \\
	Less-placeholders & 83.5\% & 99.9\% & 99.8\% \\
	More-placeholders & 74.2\% & 96.7\% & 95.8\% \\
	Generalized & 76.7\% & 74.9\% & 66.1\% \\ \midrule
	\textbf{Total} & \textbf{80.7\%} & \textbf{89.6\%} & \textbf{90.4\%} \\ \bottomrule
	\end{tabular}
\end{table}

Table~\ref{tab:Recall} shows the recall of \name{} w.r.t.\ the ground truth, i.e., the percentage of all ground truth code changes that the approach finds.
On average across the 80 queries per programming language, \name{} has a recall of 80.7\% for Java, 89.6\% for Python, and 90.4\% for JavaScript.
More specific queries tend to lead to a higher recall.
The reason is that the parse tree of a more generalized query shares fewer features with a matching code change, e.g., because a complex subtree is folded into an \code{EXPR} node.
The slightly higher recall for Python and JavaScript can be explained by two observations.
First, code changes in Java tend to be slightly larger, causing more nodes on the parse trees, which reduces the chance to find a suitable candidate change, e.g. because the probability of hash collisions is higher if there are more features.
Second, across the 80 queries, there are 236,836 ground truth code changes for Java, but only 69,626 and 59,789 for Python and JavaScript, respectively, making finding all ground truth code changes in Java a harder problem.
We discuss in Section~\ref{sec:eval parameters} that the recall can be increased even further by retrieving more candidate matches, at the expense of a slightly increased response time.

\subsection{RQ2: Efficiency and Scalability}
\label{sec:evalScalability}
A major goal of this work is to enable quickly searching through hundreds of thousands of code changes.
The following evaluates how the number of code changes to search through influences the efficiency of queries, i.e., how well \name{} scales to large amounts of changes.
As queries to run, we use the 80 queries described in \Cref{sec:eval recall}.
For each query, we measure how long \name{} takes to retrieve code changes from ten increasingly large datasets, ranging from 10,000 to 1,000,000 code changes.

\begin{figure*}
	\captionsetup[subfigure]{justification=centering}
	\centering
	\begin{subfigure}[t]{.32\linewidth}	
		\includegraphics[width=\linewidth]{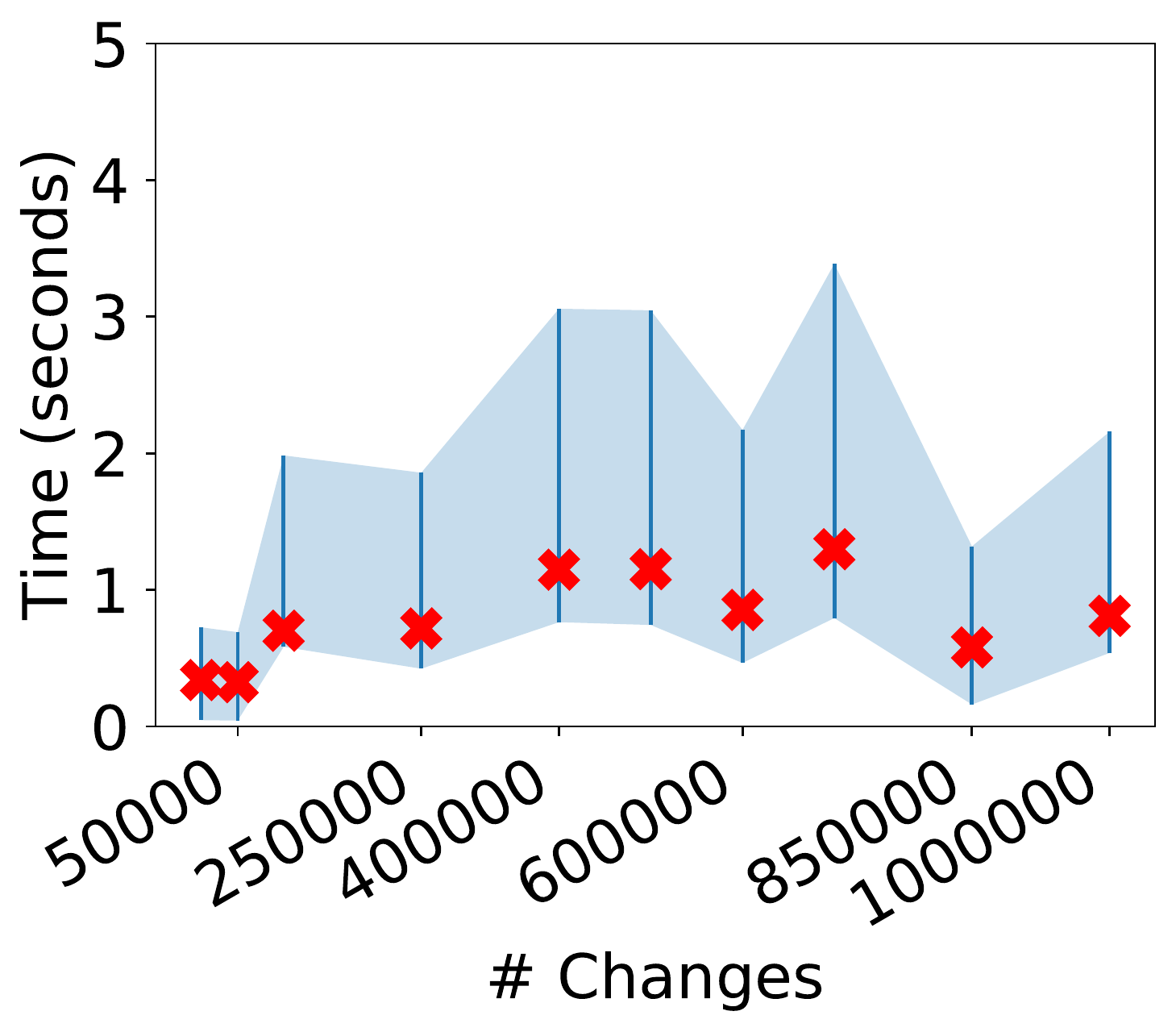}
		\caption{\name~(Java).}
		%	\label{fig:astsCodeChange}
	\end{subfigure}\hspace{0.5em}
	\begin{subfigure}[t]{.32\linewidth}
		\includegraphics[width=\linewidth]{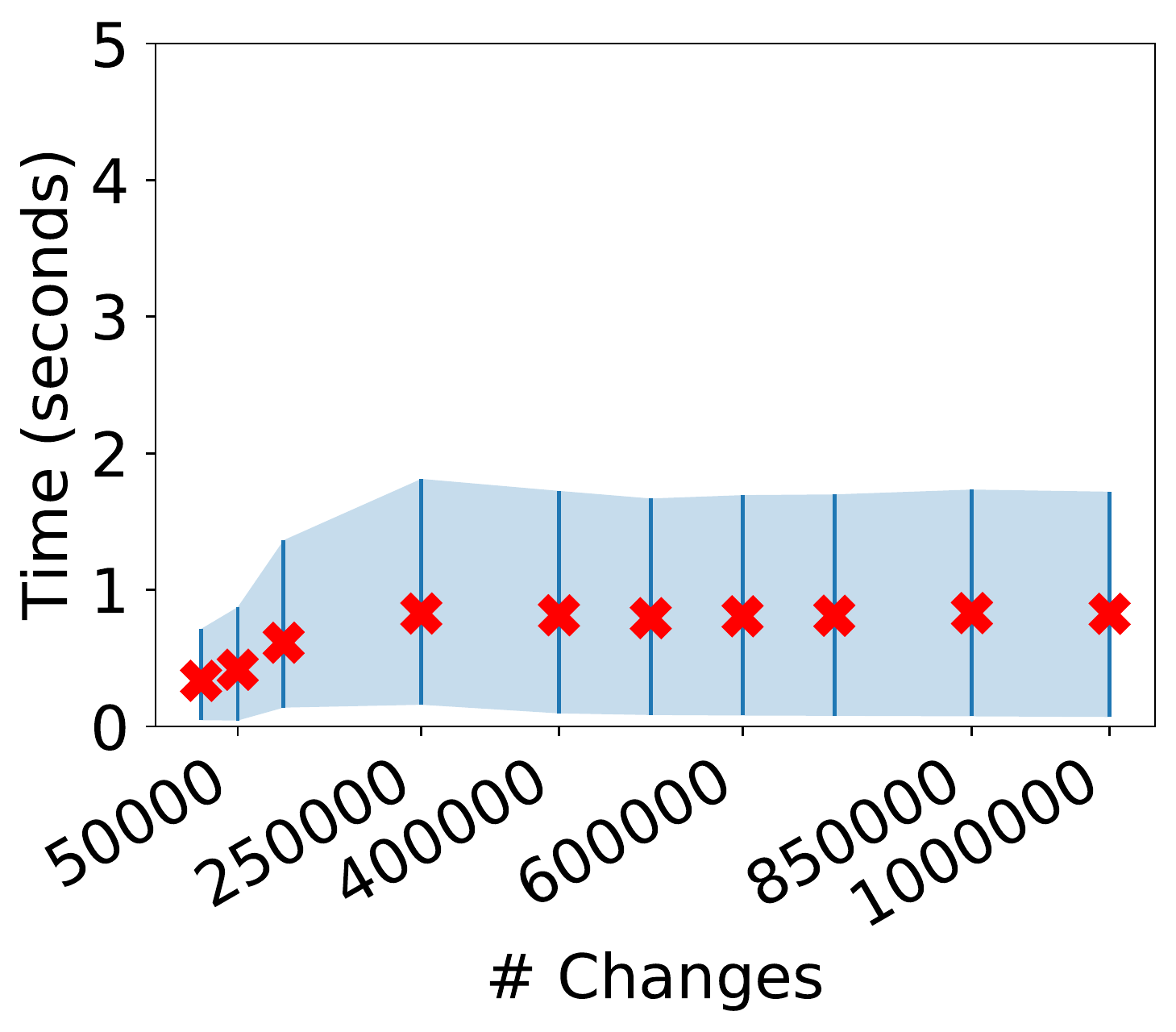}
		\caption{\name~(Python).}
	\end{subfigure}\hspace{0.5em}
\begin{subfigure}[t]{.32\linewidth}
	\includegraphics[width=\linewidth]{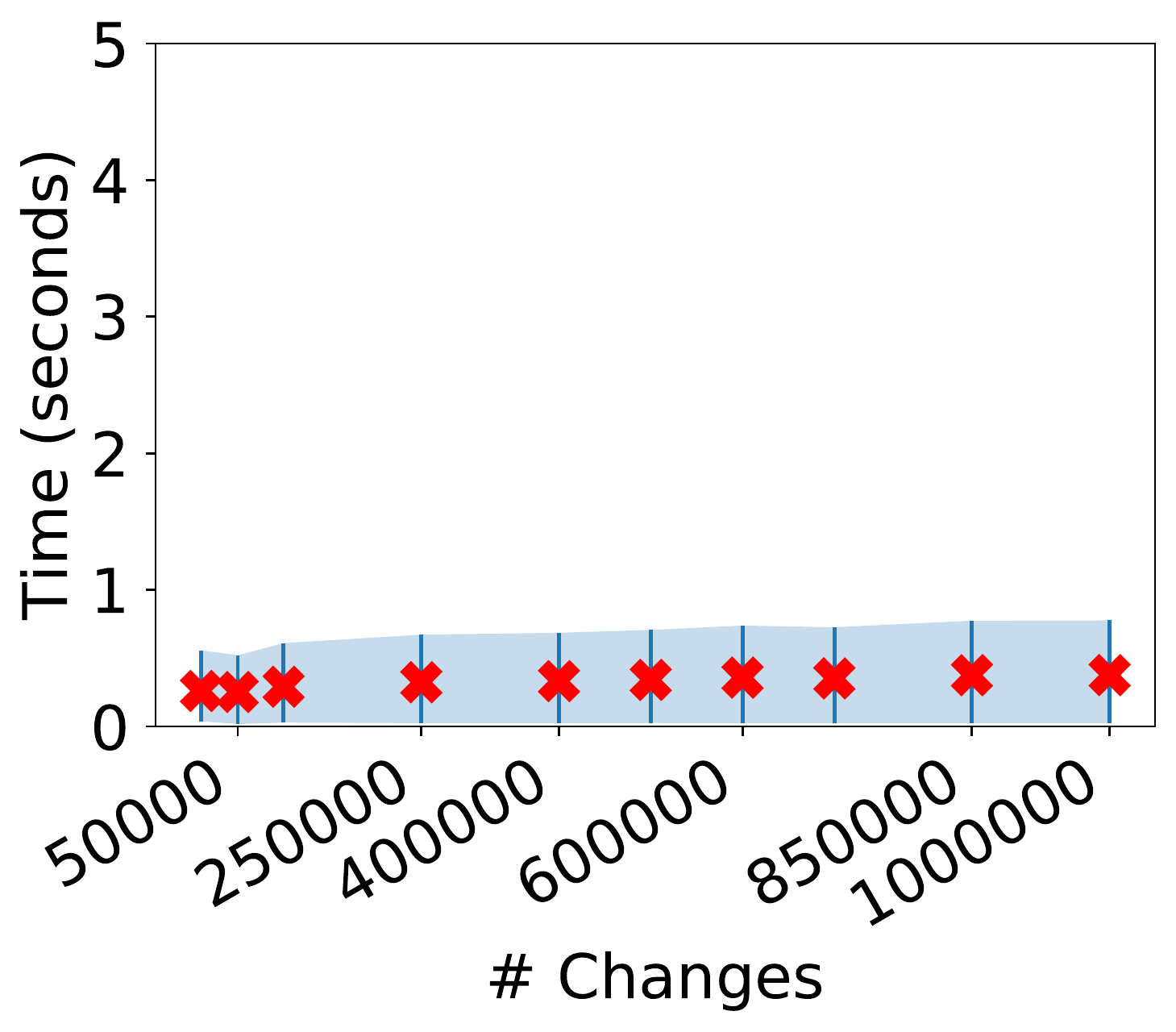}
	\caption{\name~(JavaScript).}
\end{subfigure}

\vspace{1em}
	\begin{subfigure}[t]{.32\linewidth}	
	\includegraphics[width=\linewidth]{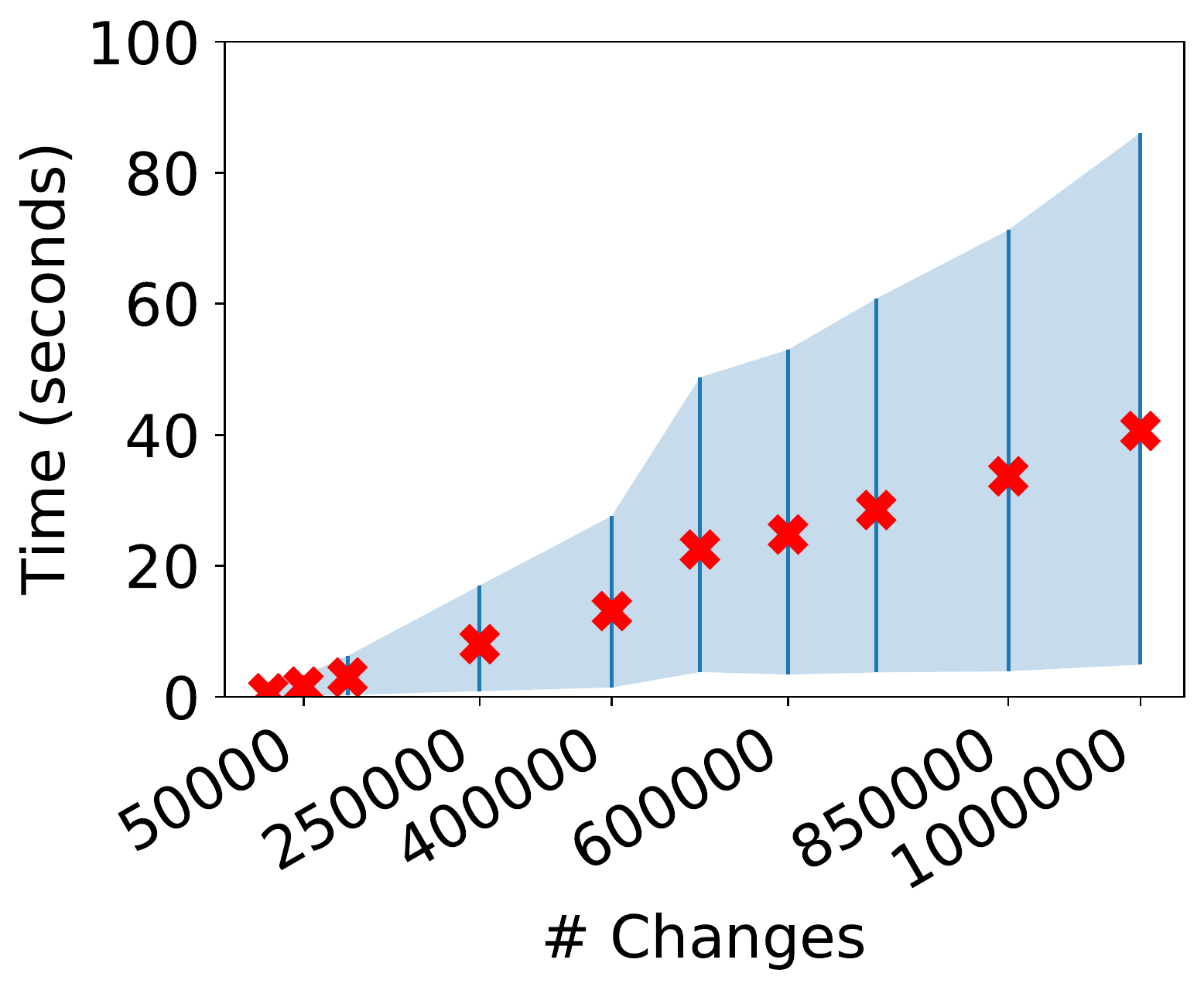}
	\caption{\name~without indexing\\(Java).}
	%	\label{fig:astsCodeChange}
\end{subfigure}
\hspace{0.5em}
\begin{subfigure}[t]{.32\linewidth}
	\includegraphics[width=\linewidth]{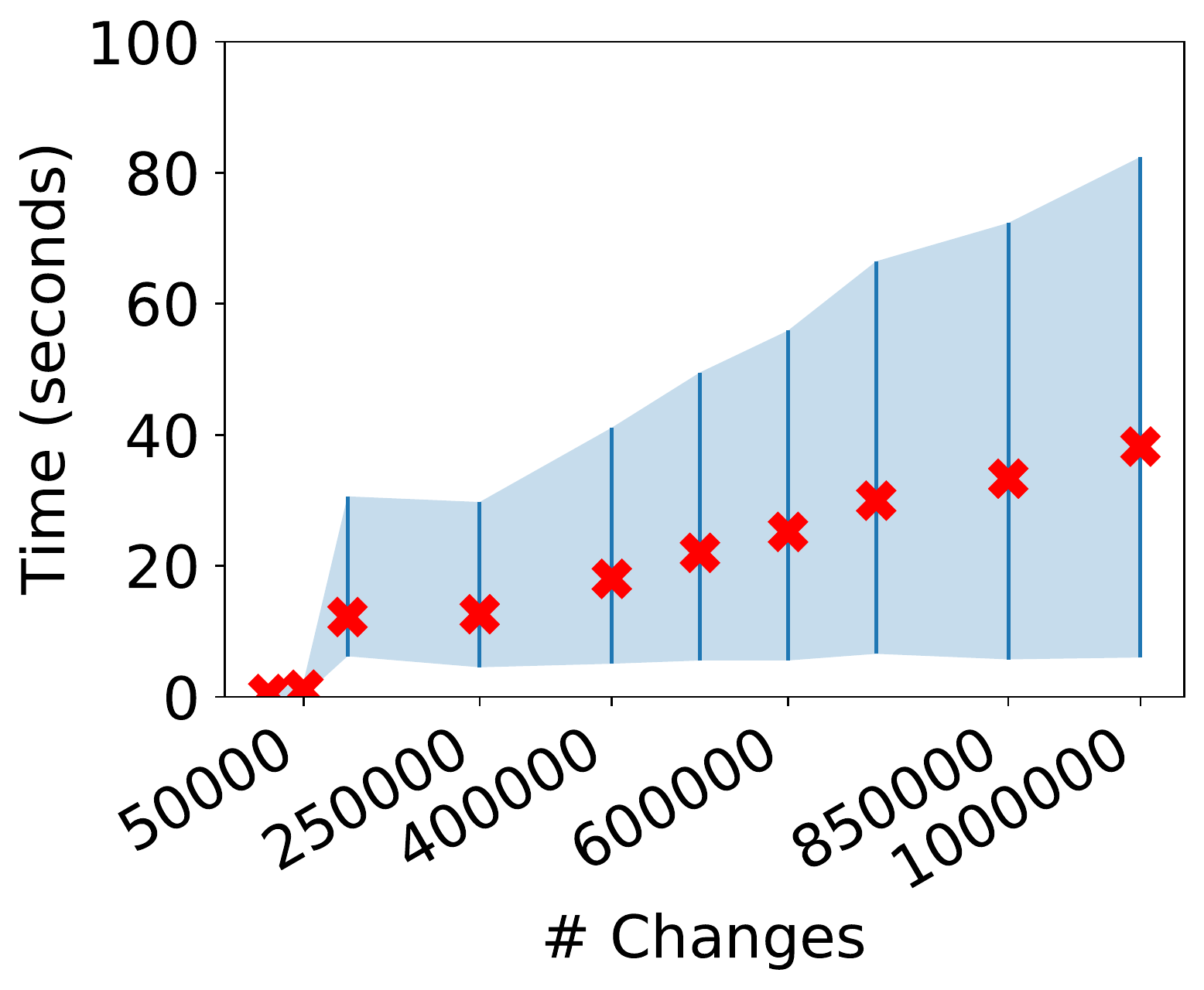}
	\centering
	\caption{\name~without indexing\\(Python).}
\end{subfigure}\hspace{0.5em}
\begin{subfigure}[t]{.32\linewidth}
	\includegraphics[width=\linewidth]{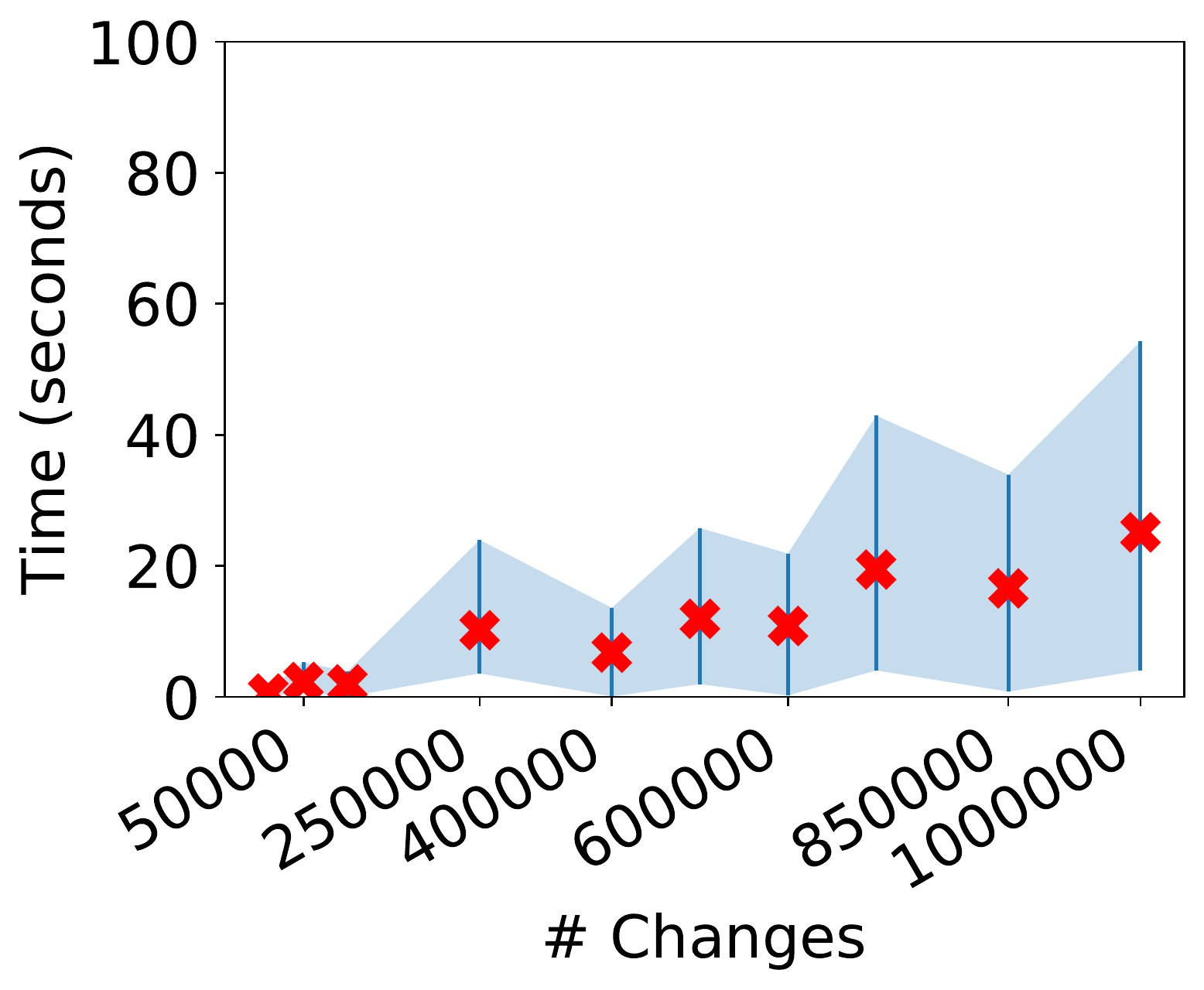}
	\caption{\name~without indexing\\(JavaScript).}
\end{subfigure}
	\caption{Response time across differently sized datasets (average and 95\% confidence interval). Top: Full \name{}. Bottom: \name{} without indexing.}
	\label{fig:scalability}
\end{figure*}

The top row of \Cref{fig:scalability} shows the results for the full \name{} approach. 
Answering a query typically takes between 0.5 and 2 seconds.
Moreover, the response time remains constant when searching through more code changes.
The reasons are (i) that FAISS~\cite{johnson2019billion} provides constant-time retrieval in the vector space, and (ii) that the time for matching candidate changes against the query is proportional to the constant number $k$ of candidate changes.
Comparing the three programming languages, we find that they yield similar performance results, which is due to the fact that most parts of our implementation are language-agnostic.
We conclude that \name{} scales well to hundreds of thousands of changes and remains efficient enough for interactive use.

The bottom row of \Cref{fig:scalability} shows the same experiment when removing the indexing and retrieval steps of \name{} (note: different y-axis).
Instead, the approach linearly goes through all code changes and compares them against a given query using the matching component only.
Answering a query takes up to 41 seconds on average, showing that the feature-based indexing is essential to ensure \name{}'s scalability.

Even though scalability is most relevant for the online part of \name{}, we also measure how long the offline part takes.
In total, analyzing a million code changes to extract feature vectors and indexing these vectors takes up to five hours.
As this is a one-time effort that does not influence the response time, we consider it acceptable in practice.

\subsection{RQ3: User Study}
\label{sec:eval user study}
%\todo{User 2 and 5 on probl 3. show positive examples that show that it is intuitive to use diffsearch}
\subsubsection{Study Setup}

We perform a user study to measure whether \name{} enables users to effectively retrieve code changes within a given time budget, and to compare our approach with a regular expression-based baseline and the GitHub Search feature.
To this end, we provide natural language descriptions of kinds of code changes and ask each user to find up to ten matching code changes per description within two minutes.
We choose this time limit based on empirical results on code search sessions, which are reported to have a median length of 89 seconds~\cite{Sadowski2015}, and to control the overall time participants of the study will have to spend.
We then ask the users how many satisfying code changes they could find.
Each user works on each kind of query with \name{}, the REGEX tool and GitHub Search.

\emph{Queries.}
The descriptions of the queries (Table~\ref{tab:user study}) are designed with two criteria in mind.
First, they cover different syntactic categories of changes, including additions (\#3, \#4, \#7), modifications (\#6), and removals (\#10) of statements; changes within existing statements (\#1, \#2, \#5, \#9); and changes that surround an existing statement with a new statement (\#8).
Second, the queries cover a diverse range of reasons for changing code, including code improvements to increase robustness (\#4, \#7, \#8), code cleanup (\#10), changes of functionality (\#6, \#9), bug fixes (\#1, \#2, \#5), and uses of a new API (\#3).
%Users search through hundreds of thousands code changes extracted from popular Java repositories.

\emph{Baselines.}
We compare \name{} against two existing tools that users might use to search for code changes.
First, we compare against a regular expression-based approach suggested in the Stack Overflow question cited in Section~\ref{sec:intro}, which we call REGEX.
Regular expressions are well known and widely used for general search tasks.
Naively applying regular expressions to the git history of many projects, as suggested on Stack Overflow, leads to unacceptably high response times (tens or even hundreds of seconds, depending on the query).
Instead, we preprocess the output of \emph{git log} by removing information unrelated to the task, such as commit messages and file names, which reduces the size of the file and makes the response time acceptable.
Second, we compare against the search feature offered by GitHub, which matches free-form queries against commits, presumably through an indexing and retrieval approach applied to the commit message and the tokens involved in a commit.
To ensure that our study participants search through the same dataset as \name{}, instead of all commits on GitHub, we create a single repository\footnote{\url{https://github.com/luca-digrazia/DatasetCommitsDiffSearch}} with all code changes in our dataset, copied from the original version histories, and then restrict GitHub's search to this repository.

\emph{Participants and setup.}
We recruit ten participants, consisting of seven PhD students, two senior undergraduate students, and one senior developer.
The participants do not overlap with the authors of this paper. % Moiz, Daniel, Jibesh, Aryaz, Cong, Patrick, Wai, Matteo, Martin, Islem (not in this order)
The user study is performed virtually with participants working from their offices or their homes.
We ask each participant to assess for each of the three tools involved in the study their level of experience (\emph{expert}, \emph{advanced}, \emph{intermediate}, or \emph{beginner}) and their usage frequency (\emph{weekly},\emph{monthly}, \emph{yearly}, or \emph{never used}).
None of the participants has previous experience with \name{}.
Regarding their experience with REGEX, four participants are \emph{advanced}, five are at \emph{intermediate} level, and one is a \emph{beginner}.
Seven participants use REGEX monthly, and three participants even weekly.
For GitHub Search, one participant is \emph{advanced}, five are \emph{intermediate}, and four are \emph{beginners}.
Two participants use it \emph{yearly}, four \emph{monthly}, three \emph{weekly}, and one has \emph{never used} it. 

The participants access \name{} through a web interface that resembles a standard search engine, but has two text input fields, for the old and new code, respectively.\footnote{The web interface is available to reviewers, see end of Section~\ref{sec:intro}.}
For REGEX, participants use a terminal and their favorite tool to search with regular expressions, e.g., \emph{grep}.
For GitHub Search we provide a link to GitHub that already restricts the search to commits in the repository created for this user study.
We provide 1,050 words of instructions to the participants, which explain the task, the query language of \name{}, how to search through raw diffs using REGEX, and GitHub Search.

\begin{table*}[]
	\centering
	\caption{Query descriptions for user study and summary of search results.}
	\label{tab:user study}
	\setlength{\tabcolsep}{1pt}
	\begin{tabular}{rlrrrrrrrrrrr}
		\hline
		\multicolumn{1}{l}{\multirow{2}{*}{\textbf{Id}}} &
		\multirow{2}{*}{\textbf{Query description}} &
		\multicolumn{11}{l}{\textbf{DiffSearch / REGEX / GitHub Search}} \\ \cline{3-6}
		\multicolumn{1}{l}{} &
		&
		\multicolumn{1}{c}{User 1} &
		\multicolumn{1}{c}{User 2} &
		\multicolumn{1}{c}{User 3} &
		\multicolumn{1}{c}{User 4} &
		\multicolumn{1}{c}{User 5} &
		\multicolumn{1}{c}{User 6} &
		\multicolumn{1}{c}{User 7} &
		\multicolumn{1}{c}{User 8} &
		\multicolumn{1}{c}{User 9} &
		\multicolumn{1}{c}{User 10} &
		\multicolumn{1}{c}{Total} \\ \hline
		1 &
		\begin{tabular}[c]{@{}l@{}}\emph{Find changes} in which a return\\ statement that returns a literal\\changes to returning the result\\of a method call.\end{tabular} &
		10/0/0 &
		10/0/0 &
		0/10/0 &
		0/0/10 &
		10/0/0 &
		7/0/0 &
		10/0/3 &
		7/0/5 &
		7/0/0 &
		\multicolumn{1}{r|}{7/0/1} &
		68/10/19 \\
		2 &
		\begin{tabular}[c]{@{}l@{}}\emph{Find changes} where the\\developer swaps the arguments\\of a method call.\end{tabular} &
		0/0/0 &
		0/0/0 &
		10/0/6 &
		10/0/1 &
		10/0/6 &
		0/0/0 &
		10/0/0 &
		10/0/0 &
		10/0/10 &
		\multicolumn{1}{r|}{10/0/0} &
		70/0/23 \\
		3 &
		\begin{tabular}[c]{@{}l@{}}\emph{Find changes} that add an import\\of a class in the form\\“import somePkg.SomeClass”.\end{tabular} &
		10/0/10 &
		0/10/10 &
		10/10/4 &
		10/10/10 &
		0/10/8 &
		10/10/10 &
		10/10/7 &
		10/0/7 &
		10/0/10 &
		\multicolumn{1}{r|}{10/0/10} &
		80/60/86 \\
		4 &
		\begin{tabular}[c]{@{}l@{}}\emph{Find changes} that add a call\\to close some resource, e.g.,\\a stream or file reader.\end{tabular} &
		0/0/3 &
		10/10/1 &
		10/10/1 &
		10/0/10 &
		10/10/1 &
		0/10/0 &
		10/10/2 &
		10/0/10 &
		10/0/0 &
		\multicolumn{1}{r|}{10/10/1} &
		80/60/29 \\
		5 &
		\begin{tabular}[c]{@{}l@{}}\emph{Find changes} where the\\condition of an if statement\\with a bodychanges from\\“-= null” to “!= null”.\end{tabular} &
		4/0/0 &
		10/0/0 &
		4/5/0 &
		0/0/10 &
		7/0/0 &
		0/0/1 &
		4/0/0 &
		4/0/0 &
		0/0/0 &
		\multicolumn{1}{r|}{5/0/0} &
		38/7/11 \\
		6 &
		\begin{tabular}[c]{@{}l@{}}\emph{Find changes} that remove\\a method call with\\one argument.\end{tabular} &
		10/0/10 &
		10/1/1 &
		10/10/10 &
		10/0/0 &
		10/0/10 &
		10/10/1 &
		10/10/0 &
		10/0/0 &
		10/0/6 &
		\multicolumn{1}{r|}{10/0/0} &
		100/31/38 \\
		7 &
		\begin{tabular}[c]{@{}l@{}}\emph{Find changes} that insert\\an assertion using\\Java’s “assert” keyword.\end{tabular} &
		10/0/6 &
		10/10/10 &
		0/10/0 &
		0/2/10 &
		10/10/2 &
		0/10/0 &
		10/10/2 &
		10/0/3 &
		10/0/0 &
		\multicolumn{1}{r|}{10/10/0} &
		70/62/33 \\
		8 &
		\begin{tabular}[c]{@{}l@{}}\emph{Find changes} in which a\\code snippet is surrounded\\with a try/catch block.\end{tabular} &
		0/0/0 &
		0/0/5 &
		0/0/1 &
		0/0/0 &
		0/0/1 &
		10/10/3 &
		4/10/4 &
		10/0/3 &
		0/0/5 &
		\multicolumn{1}{r|}{1/0/5} &
		25/0/27 \\
		9 &
		\begin{tabular}[c]{@{}l@{}}\emph{Find changes} where the\\condition of a while\\loop is changed.\end{tabular} &
		10/0/0 &
		10/10/1 &
		10/2/0 &
		10/0/0 &
		10/0/1 &
		0/0/1 &
		10/0/0 &
		0/0/0 &
		10/0/0 &
		\multicolumn{1}{r|}{10/0/0} &
		90/13/3 \\
		\multicolumn{1}{l}{10} &
		\begin{tabular}[c]{@{}l@{}}\emph{Find changes} that remove\\a call to System.out.println(...).\end{tabular} &
		10/0/6 &
		10/10/4 &
		10/10/1 &
		10/0/10 &
		10/10/5 &
		10/10/1 &
		10/10/2 &
		10/0/3 &
		10/0/1 &
		\multicolumn{1}{r|}{10/0/0} &
		100/60/33 \\ \hline
		\multicolumn{1}{l}{} &
		Total &
		64/0/35 &
		70/51/32 &
		64/67/23 &
		64/12/61 &
		60/40/34 &
		47/53/17 &
		88/50/20 &
		81/0/31 &
		77/0/32 &
		\multicolumn{1}{r|}{83/30/17} &
		711/303/302 \\ \hline
	\end{tabular}
\end{table*}

\subsubsection{Quantitative Results}

\Cref{tab:user study} shows the number of search results obtained using \name{} and REGEX.
Across the entire study, the participants find 711 code changes with \name{}, but only 303 with REGEX and 302 with GitHub Search.
Inspecting individual queries shows that, while some are harder than others, at least one user finds ten code changes for each query.
For 77.0\% of \name{} queries, users retrieve at least one code change with \name{}, whereas with REGEX, users get at least one code change for only 35.0\% of all queries, and 60.0\% of GitHub Search queries lead to at least one code change.
For 65.0\% of \name{} queries, users find the desired number of ten code changes, but only 29.0\% of users succeed with REGEX and 15.0\% with GitHub Search.
Overall, we conclude that \name{} enables users to effectively find code changes, and that the approach clearly outperforms the REGEX-based and GitHub Search baseline.

%At the end of study, we also ask each participant which approach they would prefer when searching for changes.
%All study participants decided in favor of \name{}.

\subsubsection{Qualitative Results}

To better understand the strengths and weaknesses of \name{}, we manually inspect queries formulated by users.
All the users get enough results for query \#6, e.g., with queries such as ''ID(EXPR); $\rightarrow$ \_'', underlining how easy it is querying \name{}.
Another example is query \#10, where all participants use a query similar to ''System.out.println(EXPR); $\rightarrow$ \_'', which yields 10 satisfying results.
The user study also shows how fast the participants learn to use \name{}.
For example, Users~2 and~5 on query \#3 find zero code changes with \name{}, while they find 10 code changes on query \#4 because they have learned more about the query syntax.
As another example, User~2 for query \#3 uses queries like ''\_ $\rightarrow$ import LT().LT()'' and ''\_ $\rightarrow$ import LT$<$...$>$.LT$<$...$>$'', which are syntactically invalid. After some tries the user understands the query and they perform better on the following queries.

When asking participants about their experience after the experiment, some users report difficulties in formulating precise queries on GitHub Search.
For example, for query \#6 a user says: "found many other method calls with more than one argument that were removed as well".
For query \#7 a user states: "I could find some more code that uses assert but not specifically that inserts an assert keyword".
These examples illustrate that \name{} is particularly useful when searching for non-trivial code changes and to avoid false positive results. 

%\todo{learnability?\\}
% analysis of cases where REGEX is better
While \name{} clearly outperforms REGEX and GitHub Search for all ten queries, there are some user-query pairs where REGEX and GitHub Search yields more results than \name{}.
Analyzing these cases shows two main reasons.
First, some users were effective with regular expressions by searching for simple code changes that only add or only remove a single line of code.
For example, for query \#3, some users simply searched for ``+ import (.*)''. Instead, for the same query GitHub Search has the best performance because users find precise commit messages for this kind of code change.
Second, some users formulated regular expression queries that are more general than the natural language description we provide and then manually filtered the results to find the ten relevant code changes.
For example, for query \#5, a user searched for ``\-if((.*?)){'' and then manually checked for conditions that involve \code{null}. 
Finally, Users~3 and~6 find more code changes with REGEX than the other two tools.
These users judge their REGEX experience with \emph{advanced} and \emph{intermediate}, respectively, and they both use REGEX \emph{monthly}, which they affirm to have helped them to be effective with REGEX on this task.

% limitations of REGEX
We also asked for informal feedback about the three tools, to better understand their strengths and weaknesses.
Users report three reasons for preferring \name{} over REGEX and GitHub Search.
First, they find the \name{} query more precise than regular expression syntax or free-form queries, because it builds upon the underlying programming language.
In particular, some users affirm that in two minutes they were able to type a \name{} query, but not a working regular expression, especially for complex queries, such as multi-line code changes.
Second, REGEX often was much slower than \name{} because it linearly searches through all code changes, while GitHub Search often shows commits with so many hunks that it is difficult to find a specific code change.
This inefficiency, especially for more complex code changes, caused some users to not find any relevant code changes in the given time.
Finally, some users mention that REGEX syntax is not precise enough to formulate effective queries, leading to many false positives.

%% MP: commenting out for now to not get "okay, so then do this" comment from reviewers
% limitations of DiffSearch
%The participants also pointed out ways to further improve \name{} in future work.
%Some users did not expect the retrieved changes to contain other code than what is specified in the query. However, \name{} supports partial matches (\Cref{def:match}) to enable users to find relevant code changes even when these changes are surrounded by other changed code.
%We consider preventing this problem by adding an option ``exact match'' to \name{}, which can be implemented as an extension of our matching algorithm.
%Another interesting suggestion is to add named wildcards (i.e., \code{<0>}, \code{<1>}, etc.\ in addition to \code{<...>}), e.g., to match the same block before and after the change.

\subsection{RQ4: Searching for Bug Fixes}
\label{sec:eval bug patterns}

As a case study for using \name{}, we apply it to search for instances of bug fix patterns, which could help, e.g., to establish a dataset for evaluating bug detection tools~\cite{ase2018-study}, automated program repair tools~\cite{oopsla2019}, or for training a learning-based bug detection tool~\cite{oopsla2018-DeepBugs}.
We build on a set of 16 patterns defined by prior work~\cite{Karampatsis2019a}, of which we use twelve (\Cref{tab:EffectivenessQueries}).
The remaining four bug fix patterns are all about single-token changes, e.g., changing a numeric literal or changing a modifier, which currently cannot be expressed with our query language.
For the twelve supported patterns, we formulate queries based on the descriptions of the patterns and then search for them with \name{}.
We use two different datasets for this case study.
First, a set of around 10,000 code changes, called \emph{SStuBs commits}, that contains all those commits where the prior work~\cite{Karampatsis2019a} found instances of the bug fix patterns through custom-built analysis scripts, which we call \emph{SStuBs}.
%There are some code changes that \name{} cannot parse, which we ignore for the evaluation.
Second, a set of around 1,000,000 code changes, called \emph{Large}, sampled from all the repositories analyzed in the prior work.

\begin{table}
	\caption{Effectiveness of \name{} in finding instances of bug fix patterns~\cite{Karampatsis2019a}.}
	\label{tab:EffectivenessQueries}
	\setlength{\tabcolsep}{3pt}
	\centering
	\begin{tabular}{@{}rlrrr|r@{}}
		\toprule
		& \multirow{2}{*}{Description} & \multicolumn{3}{c}{SStuBs commits (10k)} & \multicolumn{1}{c}{Large (1M)} \\ \cmidrule(l){3-6} 
		&  & \multicolumn{1}{c}{SStuBs} & \multicolumn{1}{c}{DiffSearch} & \multicolumn{1}{c|}{Both} & \multicolumn{1}{c}{DiffSearch} \\ \midrule
		1 & Change only caller & 132 & 1,880 & 121 & 5,974 \\
		2 & Change binary operator & 211 & 347 & 131 & 2,979 \\
		3 & More specific if & 130 & 592 & 116 & 5,660 \\
		4 & Less specific if & 166 & 592 & 150 & 5,387 \\
		5 & Wrong function name & 1,141 & 1,439 & 935 & 8,109 \\
		6 & Same caller, more args & 557 & 2,108 & 432 & 11,207 \\
		7 & Same caller, less args & 110 & 2,123 & 75 & 10,798 \\
		8 & Same caller, swap args & 98 & 2,285 & 89 & 9,042 \\
		9 & Change unary operator & 126 & 134 & 70 & 6,081 \\
		10 & Change binary operand & 91 & 347 & 73 & 2,136 \\
		11 & Add throws exception & 60 & 1,834 & 34 & 3,848 \\
		12 & Delete throws exception & 45 & 2,278 & 44 & 3,682 \\ \midrule
		& Total & 2,867 & 15,959 & 2,270 & 74,903 \\ \bottomrule
	\end{tabular}
\end{table}

\Cref{tab:EffectivenessQueries} shows for each bug fix pattern how many code changes the different approaches find.
\name{} returns a total of 15,959 code changes for the first dataset and 74,903 for the second dataset.
Computing the intersection with the results retrieved by SStuBs, \name{} finds 79.2\% of their changes, a result consistent with the Java recall computed in RQ1.
Moreover, \name{} finds many more matching code changes, increasing the dataset from 2,867 to 15,959 examples of bug fixes.
The reason is that our queries are more general than the custom analysis scripts in SStuBs and include, e.g., also code changes that perform other changes besides the specific bug fix.
The number of code changes found by \name{} is higher than the number of commits (10k) because a single commit may match multiple patterns.
For example, a change that swaps two arguments and modifies a function name will appear in patterns~5 and~8.
Overall, \name{} is effective at finding various examples of bug fix patterns, showing the usefulness of the approach for creating large-scale datasets.

\subsection{RQ5: Impact of Parameters}
\label{sec:eval parameters}
We perform a sensitivity analysis for the two main parameters of \name{}:
the length $l$ of feature vectors (Section~\ref{sec:features}), and the number $k$ of candidate matches retrieved via the feature vectors (Section~\ref{sec:indexingRetrieval}).
We select a set of values from 1,000 to 20,000 for $k$ and from 500 to 4,000 for $l$, i.e., values below and above the defaults, and then measure their impact on the time to answer queries, the recall, and the size of the index.

\begin{table}[t]
	\caption{Impact of length $l$ of feature vectors and number $k$ of candidates (default configuration is bold).}
	\label{tab:parameters}
	\setlength{\tabcolsep}{9pt}
	\centering
\begin{tabular}{@{}rrrrrrr@{}}
	\toprule
	$k$ & $l$ & \multicolumn{3}{@{}c@{}}{Response time (s)} & Recall & Size of \\
	\cmidrule{3-5}
	&& min & avg & max & (\%) & index (GB) \\
	\midrule
	\multicolumn{7}{@{}l@{}}{\emph{Java:}}   \\
	\midrule
	1,000 & 1,000 & 1.5 & 1.9 & 3.5 & 71.8 & 4.0 \\
	\textbf{5,000} & \textbf{1,000} & \textbf{1.5} & \textbf{2.2} & \textbf{9.0} & \textbf{80.7} & \textbf{4.0} \\
	10,000 & 1,000 & 1.7 & 2.5 & 9.4 & 84.9 & 4.0 \\
	20,000 & 1,000 & 1.8 & 3.1 & 17.7 & 87.3 & 4.0 \\
	5,000 & 500 & 0.8 & 1.3 & 8.1 & 79.3 & 2.0 \\
	5,000 & 2,000 & 3.0 & 4.2 & 9.9 & 80.6 & 8.0 \\
	5,000 & 4,000 & 5.8 & 7.4 & 15.3 & 78.1 & 16.0 \\
	\midrule
	\multicolumn{7}{@{}l@{}}{\emph{Python:}}   \\
	\midrule
	1,000 & 1,000 & 3.0 & 4.1 & 5.5 & 81.9 & 4.1 \\
	\textbf{5,000} & \textbf{1,000} & \textbf{1.8} & \textbf{2.4} & \textbf{3.5} & \textbf{89.8} & \textbf{4.1} \\
	10,000 & 1,000 & 3.5 & 5.0 & 8.9 & 91.6 & 4.1 \\
	20,000 & 1,000 & 4.1 & 6.0 & 12.4 & 93.7 & 4.1 \\
	5,000 & 500 & 1.0 & 1.6 & 3.1 & 86.6 & 2.0 \\
	5,000 & 2,000 & 2.7 & 4.9 & 40.8 & 89.8 & 8.1\\
	5,000 & 4,000 & 6.1 & 7.9 & 13.1 & 83.4 & 16.3 \\
	\midrule
	\multicolumn{7}{@{}l@{}}{\emph{JavaScript:}} \\
	\midrule
	1,000 & 1,000 & 1.2 & 1.9 & 2.8 & 85.4 & 4.0 \\
	\textbf{5,000} & \textbf{1,000} & \textbf{1.3} & \textbf{2.0} & \textbf{2.8} & \textbf{90.4} & \textbf{4.0} \\
	10,000 & 1,000 & 1.4 & 2.3 & 3.3 & 94.0 & 4.0 \\
	20,000 & 1,000 & 1.8 & 2.9 & 5.7 & 95.6 & 4.0 \\
	5,000 & 500 & 0.7 & 1.2 & 2.1 & 90.3 & 2.0  \\
	5,000 & 2,000 & 3.1 & 4.5 & 5.4 & 92.5 & 8.0 \\
	5,000 & 4,000 & 5.1 & 9.2 & 12.8 & 88.6 & 16.1 \\
	\bottomrule
\end{tabular}
\end{table}

Table~\ref{tab:parameters} shows the results.
We find that retrieving more candidate code changes, i.e., a higher $k$, slightly increases the response time.
The reason is that matching more code changes against the query increases the time taken by the matching phase.
On the positive side, increasing $k$ increases the recall, reaching 87.3\% for Java, 93.7\% for Python, and 95.6\% for JavaScript  when $k$=20,000, while still providing an acceptable average response time.
Parameter $l$ increases the time to answer a query because a larger feature vector slows down the  nearest neighbor search.
Likewise, a larger $l$ also increases the size of the index.
Since increasing $l$ beyond our default does not significantly increase recall, we use $l$=1,000 as the default to have a manageable index size and a reasonable response time. As a result, users can adjust the parameters based on their usage scenario. They can use a higher $k$ if they prefer recall over efficiency, or a lower $k$ if they prefer the opposite.

%We also notice that a higher parameter $l$ gives worse results in terms of recall. We analyze the feature vectors and we find that, e.g., for JavaScript, the index with $l$=1,000 contains 2.8\% of 1s and 1.4\% for the index with $l$=4,000. As a result, a sparse vector makes the Nearest neighbor search less effective. We notice a similar behavior for Java and Python.

\subsection{RQ6: Queries vs.\ Search Results}
\label{sec:eval query vs results}

The goal of performing a search is to obtain more information than provided in the query.
To assess to what extent \name{} serves this purpose by characterizing queries and the resulting search results in two ways.
These experiments are done on all three currently supported languages, using the 80 queries described in RQ1.

First, we quantify the number of results obtained via a single query. We compute the average number of code changes retrieved by \name{} among the 80 queries.  
We find an average of 646 results for Java, 269 for Python, and 280 for JavaScript. As a result, we can conclude that typing a single \name{} query results in a significant amount of information retrieved.

Second, we approximate the amount of information in a query and the resulting search results by counting the number of characters they are composed of. For the results with multiple code changes we compute the average of their size.
We find an average query size of 95 and an average result size of 136 for Java, an average query size of 47 and an average result size of 67 for Python, and an average query size of 34 and an average result size of 55 for JavaScript.
As a result, we can conclude that the result of \name{} queries contains 29.9\%, 29.8\%, and 38.2\% more information than provided in the query for Java, Python and JavaScript, respectively.

In conclusion, we show that the effort to type a \name{} query has benefits in the quantity of information retrieved.

\section{Limitations and Future Work}

Our approach has some limitations that will be interesting to address in future work.
% Parsing and ambiguity
First, in Section~\ref{sec:trees}, we explain the challenge of parsing incomplete parse trees.
We extend the ANTLR4 grammar for the target programming languages with optional rules to parse incomplete snippets of code that commonly occur in hunks.
These extensions cover most but not all hunks, and we plan to enable parsing of an even larger range of incomplete code snippets in the future.
Second, our approach of parsing individual hunks will be non-trivial to apply to languages that make heavy use of macros, such as C.
The reason is that, when trying to parse a single hunk, the definitions of macros are not available.
%
% Neural feature
Finally, Section~\ref{sec:features} describes the features we design for code changes. Future work could either extend those features with other features or apply neural networks that learn to map code changes into continuous vector representations, such as CC2Vec~\cite{Hoang2020} and Commit2Vec~\cite{cabrera2021commit2vec}.
\section{Related Work}
\emph{Code Search.}
Code search engines allow users to find code snippets based on method signatures~\cite{reissCodeSearch}, existing code examples~\cite{kim2018facoy,Luan2019,Premtoon2020}, or natural language queries~\cite{Gu2018,Sachdev2018,cambronero2019deep}.
Sourcerer provides an infrastructure that combines several of the above ideas~\cite{sourcerer}.
Early work by Paul et al.~\cite{Paul1994} proposes a mechanism similar to the placeholders in our query language.
The most important difference between these approaches and \name{} is that we search for changes of code, not for code snippets within a single snapshot of code.
Another difference is that \name{} guarantees that all search results match the given query, whereas the existing techniques, with the exception of \cite{Premtoon2020}, are aimed at similarity only.

Prequel has a goal similar to \name{}, and matches C, and partially also C++ and Java, patches against user-provided rules that the code before and after a patch must comply with~\cite{Lawall2016prequel}.
The approaches differ in four aspects.
First, Prequel can describe all the parts of a commit, including the code surrounding and relationship between multiple hunks. Instead, \name{} focus on single hunks. 
Second, Prequel's rules are based on the semantic patch language of Coccinelle~\cite{Lawall2018} and may include executable code, e.g., queries are Turing-complete.
In contrast, our queries are purely declarative and build on the underlying programming language.
Third, Prequel pre-filters commits based on a regular expression or indexing, followed by a linear search through all remaining commits.
As a result, answering a query may take minutes or, if the pre-filtering is not effective, even longer~\cite{Lawall2016prequel}.
In contrast, \name{} avoids a linear search via feature-based retrieval, and hence, responds to queries across hundreds of thousands of code changes within seconds. The indexing module could be used also in Prequel in principle.
Finally, Prequel is evaluated on a different problem than \name{}: obtaining examples to motivate device driver porting from 300,000 commits of the Linux kernel.

Several ideas to improve the user's interaction with a code search engine have been proposed, such as refining search results based on user feedback about the quality of results~\cite{martie2017understanding,sivaraman2019active}.
Other work resolves vocabulary mismatches between queries and code~\cite{sirres2018augmenting}. Finally, code2vec~\cite{Alon2019} represents a code snippet with a single-size feature vector, but with respect to \name{} they use ASTs and neural networks. The main difference with our work is that code2vec converts snippets of code in vectors, instead \name{} converts code changes in vectors.
Future work could adopt similar ideas to searching for code changes.

\emph{Code Changes as Edit Scripts.}
To reason about code changes, several techniques derive edit scripts on ASTs~\cite{Fluri2007,Hashimoto2008,Falleri2014,Erdweg2021}, providing an abstract description of the change that can then be applied elsewhere~\cite{Meng2011}.
Lase generalizes from multiple code changes into a single edit script~\cite{Meng2013}.
Future work could explore using an edit script-based representation of code changes to search for code changes.
An advantage of our parse tree-based feature extraction is that it does not require aligning the old and new code, allowing us to featurize hundreds of thousands of code changes in reasonable time. % Yes, this statement is still okay, because we discarded that features.

\emph{Mining Code Changes.}
Work on mining code repositories and learning from code changes shows development histories to be a rich source of implicitly stored knowledge.
For example, existing approaches leverage version histories to
extract repetitive code changes~\cite{negara2014mining,Nguyen2019,nguyen2013study},
predict code changes~\cite{Tufano2019},
predict bugs~\cite{Livshits2005,Kim2008}, or to 
learn about API usages~\cite{nguyen2016api,Paletov2018}.
Mining approaches typically consider all code changes in a project's version history or filter changes using simple patterns, e.g., keywords in commit messages.
In contrast, \name{} allows for identifying code changes that match a specific query.

\emph{Learning from Code Changes.}
Large sets of code changes enable learning-based techniques.
One line of work learns from specific kinds of changes, e.g., fixes of particular bug patterns, how to apply this kind of change to other code for automated program repair~\cite{Rolim2017,Rolim2018,oopsla2019}.
Another line of work ranks potential program repairs based on their similarity to common code change patterns~\cite{Le2016}.
\name{} could help gather datasets of changes for these approaches to learn from, e.g., based on queries for bug fixing patterns.

\emph{Representation Learning on Commits.}
The feature extractor of \name{} relates to techniques for learning vector representations of commits, such as CC2Vec~\cite{Hoang2020} and Commit2Vec~\cite{cabrera2021commit2vec}.
These techniques train a model on some ``pseudo task'' for which abundant training data is easily available, e.g., predicting the words in the commit message~\cite{Hoang2020} or whether a commit is labeled as security-critical~\cite{cabrera2021commit2vec}.
Once trained, the vector representations produced by a representation learning model could, in principle, be used as an alternative to the feature vectors of \name{}.
In practice, integrating CC2Vec and Commit2Vec into our approach is non-trivial because both approaches focus on entire commits, which may include many hunks distributed across multiple files, whereas \name{} retrieves code changes at hunk-level granularity.
Finding an appropriate pseudo task for representation learning on individual hunks, and integrating the resulting embeddings into \name{}, could be interesting future work.

%CC2Vec~\cite{Hoang2020} implement a neural network approach to extract features from code changes that is evaluated on three different tasks: log message generation (on Java code), bug fix patch finder (on C code) and just-in-time defect prediction (on C++ and Python code).
%Commit2Vec uses transfer learning to extract code changes features for the classification of security-relevant commits.
%The main difference with respect \name{} is that they extract features from an entire commit and not on single code changes.
% The main difference with respect to DiffSearch’s feature extraction is that CC2Vec granularity is not based on a single hunk, but it splits the code changes per file and then it computes the feature. Moreover, they include the log message of each commit as input of the pre-trained model, that is usually a generic summary of the changes that is not specific for one specific hunk, making this approach less relevant for single hunk code changes that is our target.

\emph{Other Analyses of Code Changes.}
There are various other analyses of code changes, of which we discuss only a subset here.
Hashimoto et al.\ propose a technique for reducing a diff to the essence of a bug~\cite{Hashimoto2018}.
Nielsen et al.~\cite{nielsen2021semantic} use JavaScript code change templates to fix code broken due to library evolution. 
Another approach automatically documents code changes with a natural language description~\cite{Buse2010}.
SCC~\cite{giger2011comparing} and DeepJIT~\cite{Hoang2019} are predictive models that estimate how likely a code change is to introduce a bug.
A related problem is to find the bug-inducing code change for a given bug report~\cite{Wen2016,Wu2018}.
DiffBase~\cite{Wu2021} encodes facts about different versions of a program to facilitate multi-version program analyses.
CodeShovel~\cite{Grund2021} tracks a method from its creation to its current state throughout a version history. All these approaches relate to our work by also reasoning about code changes, but they aim for different goals than \name{}.

\emph{Clone Detection.}
\name{} relates to code clone detectors~\cite{kamiya2002ccfinder,Li2006,jiang2007deckard,roy2008nicad,sajnani2016sourcerercc}, as answering a query resembles finding clones of the query.
In particular, \name{} compares a query against code changes in a way similar to Type-1 clones, and when using placeholders in the query, similar to Type-2 and Type-3 clones.
Clone detectors are typically evaluated on a single snapshot of a code base, and they may take several minutes or even hours to terminate~\cite{sajnani2016sourcerercc}.
In principle, one could use an off-the-shelf code clone detector to search for specific kinds of code changes, where the old and new parts of the query must be clones of the old and new parts of a change, respectively.
However, this approach would search for clones among all code changes for each query, which may not be fast enough for an interactive search engine.
Some clone detectors summarize code in ways related to our feature extraction.
For example, Deckard~\cite{jiang2007deckard} computes characteristic vectors of parse trees and SourcererCC~\cite{sajnani2016sourcerercc} indexes large amounts of code into a bag-of-tokens representation.
Integrating such ideas into the feature-based retrieval in \name{} could further improve recall.
Inoue et al.~\cite{Inoue2020} propose a code clone detector that supports special tokens, such as $\$$, $*$, $\#$, to express exact matching, repetitions, and more, similar to regular expressions.
However, their approach cannot express relationships between an old and a new code snippet, as supported by \name{}.
Nguyen et al.~\cite{Nguyen2017exploring} perform an empirical study on a large dataset of Java and C\# using API2VEC based on Word2Vec to create feature vectors from APIs. They find this kind of representation successful, because APIs with similar usage context have closer feature vectors using this representation. \name{} differs from their approach because match code changes and because perform a matching based on the syntax of the code more than their usage context.

\section{Conclusion}

We present a scalable and precise search engine for code changes.
Given a query that describes code before and after a change, the approach retrieves within seconds relevant examples from a corpus of a million code changes.
Our query language extends the underlying programming language with wildcards and placeholders, providing an intuitive way of formulating queries to search for code changes.
Key to the scalability of \name{} is to encode both queries and code changes into a common feature space, enabling efficient retrieval of candidate search results.
Matching these candidates against the query guarantees that every returned search result indeed fits the query.
The approach is mostly language-agnostic, and we empirically evaluate it on Java, JavaScript, and Python.
\name{} answers most queries in less than a second, even when searching through large datasets.
The recall ranges between 80.7\% and 90.4\%, depending on the target language, and can be further increased at the expense of response time.
We also show that users find relevant code changes more effectively with \name{} than with a regular expression-based search and GitHub Search.
Finally, as an example of how the approach could help researchers, we use it to gather a dataset of 74,903 code changes that match recurring bug fix patterns.
We envision \name{} to serve as a tool useful to both practitioners and researchers, and to provide a basis for future work on searching for code changes.

\section*{Acknowledgment}
This work was supported by the European Research Council (ERC, grant agreement 851895), and by the German Research Foundation within the ConcSys and DeMoCo projects.

\IEEEdisplaynontitleabstractindextext

\IEEEpeerreviewmaketitle

%\balance
\bibliographystyle{IEEEtran}
\bibliography{referencesMP,moreReferences}

%\enlargethispage{-1in}
\begin{IEEEbiography}[{\includegraphics[width=1in,height=1.25in,clip,keepaspectratio]{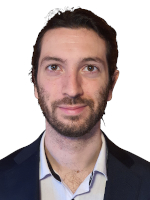}}]{Luca Di Grazia}
is a PhD student in the Department of Computer Science at the University of Stuttgart in Germany, advised by Michael Pradel. His main research interests are code evolution and maintenance, mining software repositories, and program analysis.
\end{IEEEbiography}

\begin{IEEEbiography}[{\includegraphics[width=1in,height=1.25in,clip,keepaspectratio]{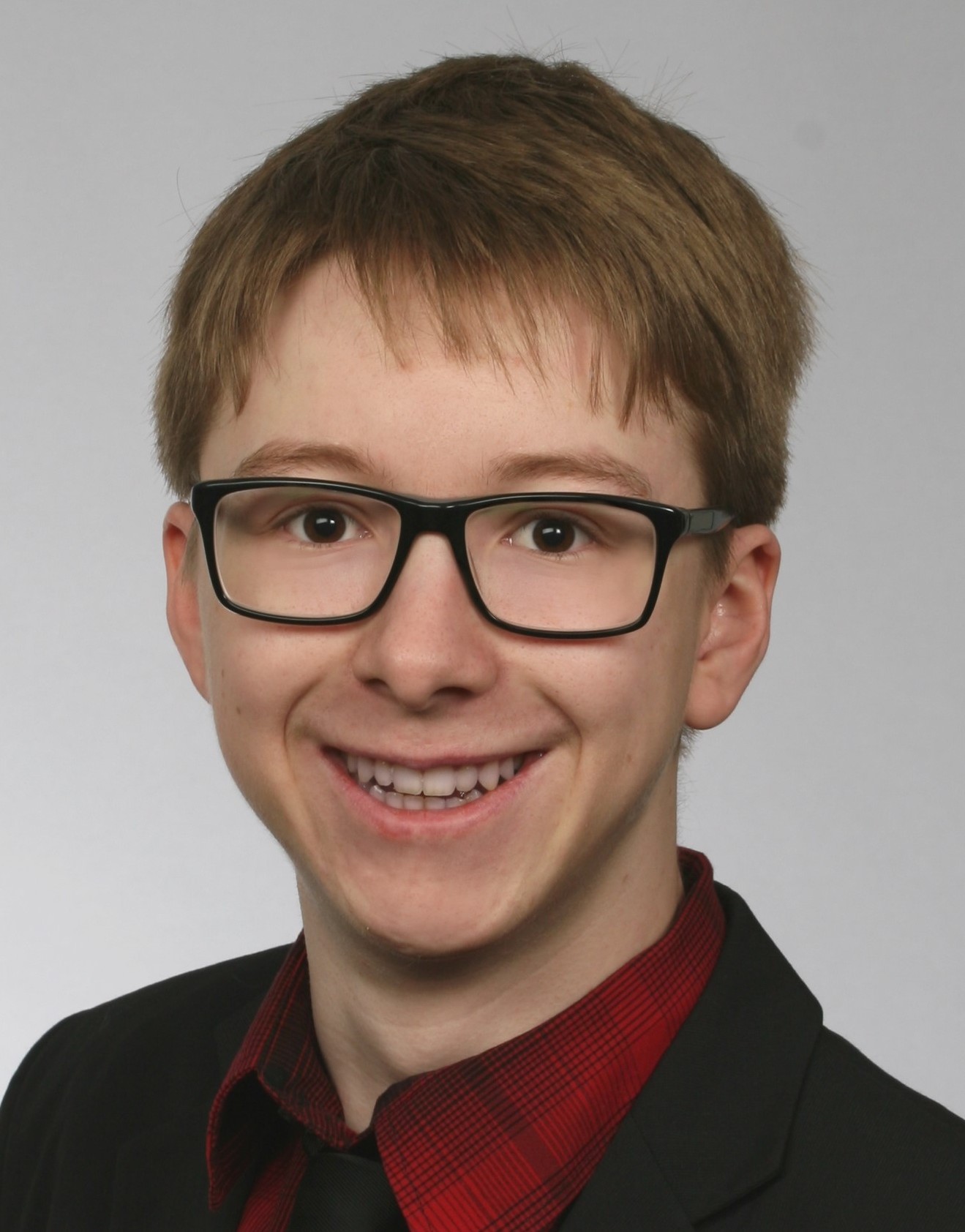}}]{Paul Bredl}
is a master student in Software Engineering at the University of Stuttgart. During and after his bachelor studies he worked as a software developer, maintaining and extending enterprise systems.

\end{IEEEbiography}

\begin{IEEEbiography}[{\includegraphics[width=1in,height=1.25in,clip,keepaspectratio]{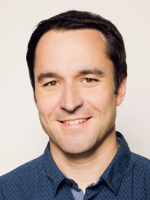}}]{Michael Pradel}
is a full professor at the University of Stuttgart. His research interests span software engineering, programming languages, security, and machine learning, with a focus on tools and techniques for building reliable, efficient, and secure software.
\end{IEEEbiography}\vfill

\end{document}